\documentclass{article}

\usepackage[scr=boondoxo,scrscaled=1.05]{mathalfa}
\usepackage[pdftex]{graphicx}
\usepackage{amsmath,amssymb,hhline,tikz,mathtools}
\usepackage{enumerate}
\usetikzlibrary{shapes.geometric,decorations.markings}
\usetikzlibrary{calc}
\usepackage[title]{appendix}
\usepackage{etoolbox}
\patchcmd{\appendices}{\quad}{: }{}{}
\usepackage{algorithm}
\usepackage{algorithmicx}
\usepackage{algpseudocode}
\usepackage{rotating}
\usepackage{afterpage}
\usepackage{xstring}
\usepackage{xcolor}
\usepackage{cmap}
\usepackage{mathdots}
\usepackage{colortbl}
\usepackage{booktabs}
\usepackage{textcomp}     
\usetikzlibrary{backgrounds}
\usepackage{pgfplots}
\pgfplotsset{compat=1.12}
\pgfmathsetseed{5112}
\usepackage{multirow}
\usepackage{pgfplotstable}
\usepackage{adjustbox}
\usepackage{diagbox}

\PassOptionsToPackage{hyphens}{url}
\usepackage{hyperref}
\hypersetup{
    pdftitle={Word Embedding Techniques for Malware Classification},
    pdfauthor={Mark Stamp},
    bookmarksnumbered=true,     
    bookmarksopen=true,         
    bookmarksopenlevel=3,       
    colorlinks=true,linkcolor=black,citecolor=black,urlcolor=blue,filecolor=blue,
    pdfstartview=Fit,           
    pdfpagemode=UseOutlines,
    pdfpagelayout=SinglePage
}

\definecolor{darkgreen}{rgb}{0.125,0.5,0.169}

\algrenewcommand{\algorithmiccomment}[1]{{\color{red}{\tt //}\ #1}}
\algnewcommand{\Initialize}[1]{%
  \State \textbf{Initialize:}
  \Statex \hspace*{\algorithmicindent}\parbox[t]{.8\linewidth}{\raggedright #1}
}
\algnewcommand{\Given}[1]{%
  \State \textbf{Given:}
  \Statex \hspace*{\algorithmicindent}\parbox[t]{.8\linewidth}{\raggedright #1}
}

\long\def\symbolfootnotetext[#1]#2{\begingroup%
\def\thefootnote{\fnsymbol{footnote}}\footnotetext[#1]{#2}\endgroup}

\let\oldsqrt\sqrt
\def\sqrt{\mathpalette\DHLhksqrt}
\def\DHLhksqrt#1#2{%
\setbox0=\hbox{$#1\oldsqrt{#2\,}$}\dimen0=\ht0
\advance\dimen0-0.2\ht0
\setbox2=\hbox{\vrule height\ht0 depth -\dimen0}%
{\box0\lower0.4pt\box2}}

\def\clap#1{\hbox to 0pt{\hss#1\hss}}

\allowdisplaybreaks

\def\figureFastSpeed{s}\def\figureSpeed{f}
\let\figureFastSpeed=\figureSpeed
\def\selectFigureSpeed#1#2{
\if\figureSpeed\figureFastSpeed #1\else #2\fi}

\def\srowvecc#1#2{(\!\begin{array}{cc} 
      \noexpandarg\IfBeginWith{#1}{-}{\! #1}{#1}
    & #2\kern-0.5pt\end{array}\!)}
\def\rowvecc#1#2{\left(\!\begin{array}{cc} 
      \noexpandarg\IfBeginWith{#1}{-}{\! #1}{#1}
    & #2\kern-0.5pt\end{array}\!\right)}
\def\rowveccc#1#2#3{\left(\!\begin{array}{ccc} 
      \noexpandarg\IfBeginWith{#1}{-}{\! #1}{#1}
    & #2 
    & #3\kern-0.5pt\end{array}\!\right)}
\def\rowvecccc#1#2#3#4{\left(\!\begin{array}{cccc}
      \noexpandarg\IfBeginWith{#1}{-}{\! #1}{#1}
    & #2 
    & #3 
    & #4\kern-0.5pt\end{array}\!\right)}
\def\srowvecccc#1#2#3#4{\bigl(\!\begin{array}{cccc}
      \noexpandarg\IfBeginWith{#1}{-}{\! #1}{#1}
    & #2 
    & #3 
    & #4\kern-0.5pt\end{array}\!\bigr)}
\def\rowveccccc#1#2#3#4#5{\left(\!\begin{array}{ccccc} 
      \noexpandarg\IfBeginWith{#1}{-}{\! #1}{#1}
    & #2
    & #3
    & #4
    & #5\kern-0.5pt\end{array}\!\right)}
\def\srowvecccccc#1#2#3#4#5#6{(\!\begin{array}{cccccc} 
      \noexpandarg\IfBeginWith{#1}{-}{\! #1}{#1}
    & #2
    & #3
    & #4
    & #5
    & #6\kern-0.5pt\end{array}\!)}
\def\rowvecccccc#1#2#3#4#5#6{\left(\!\begin{array}{cccccc} 
      \noexpandarg\IfBeginWith{#1}{-}{\! #1}{#1}
    & #2
    & #3
    & #4
    & #5
    & #6\kern-0.5pt\end{array}\!\right)}

\def\figureType{*}\def\figureSlowType{slowType}
\def\selectFigureType#1#2{
\if\figureType\figureSlowType #1\else #2\fi}
\makeatletter
\newcommand{\lowsub}[1]{\mathpalette{\raisem@th{#1}}}
\newcommand{\raisem@th}[3]{\raisebox{-#1}{$#2#3$}}
\makeatother
\def\halfthin{\kern 0.083em}

\DeclareMathOperator{\thth}{th}

\pgfkeys{
    /pgf/number format/precision=2, 
    /pgf/number format/fixed zerofill=true }
    
\pgfplotstableset{
    /color cells/min/.initial=0,
    /color cells/max/.initial=1000,
    /color cells/textcolor/.initial=,
    color cells/.code={%
        \pgfqkeys{/color cells}{#1}%
        \pgfkeysalso{%
            postproc cell content/.code={%
                \begingroup
                \pgfkeysgetvalue{/pgfplots/table/@preprocessed cell content}\value
\ifx\value\empty
\endgroup
\else
                \pgfmathfloatparsenumber{\value}%
                \pgfmathfloattofixed{\pgfmathresult}%
                \let\value=\pgfmathresult
                \pgfplotscolormapaccess
                    [\pgfkeysvalueof{/color cells/min}:\pgfkeysvalueof{/color cells/max}]%
                    {\value}%
                    {\pgfkeysvalueof{/pgfplots/colormap name}}%
                \pgfkeysgetvalue{/pgfplots/table/@cell content}\typesetvalue
                \pgfkeysgetvalue{/color cells/textcolor}\textcolorvalue
                \toks0=\expandafter{\typesetvalue}%
                \xdef\temp{%
                    \noexpand\pgfkeysalso{%
                        @cell content={%
                            \noexpand\cellcolor[rgb]{\pgfmathresult}%
                            \noexpand\definecolor{mapped color}{rgb}{\pgfmathresult}%
                            \ifx\textcolorvalue\empty
                            \else
                                \noexpand\color{\textcolorvalue}%
                            \fi
                            \the\toks0 %
                        }%
                    }%
                }%
                \endgroup
                \temp
\fi
            }%
        }%
    }
}

\makeatletter
\newcommand*\bigcdot{\mathpalette\bigcdot@{.5}}
\newcommand*\bigcdot@[2]{\mathbin{\vcenter{\hbox{\scalebox{#2}{$\m@th#1\bullet$}}}}}
\makeatother

\def\O{{\cal O}}

\def\k{\kern 2.75pt}

\newlength{\xxxxx}
\def\sscoin{%
  \leavevmode
  \vtop{\offinterlineskip 
    \setbox0=\hbox{\scriptsize S}%
    \setbox2=\hbox to\wd0{\hfil\hskip-.03em
    \vrule height .3ex width .08ex\hskip .08em
    \vrule height .3ex width .08ex\hfil}
    \vbox{\copy2\box0}\box2}}
\newcommand\affil[2]{%
  \begingroup
  \renewcommand\thefootnote{}\footnote{\llap{$\hbox{}^{#1}\hbox{}$}#2}%
  \addtocounter{footnote}{-1}%
  \endgroup
}
\newcommand\markonly[1]{%
$\hbox{}^{\mbox{\kern4.5pt,\kern0.75pt #1}}$
}

\advance\oddsidemargin by -0.525in
\advance\textwidth by 1.05in
\advance\topmargin by -0.625in
\advance\textheight by 1.25in

\title{\vspace{-0.5in}Word Embedding Techniques for Malware Evolution Detection}

\author{
Sunhera Paul\thanks{sunherabarunkumar.paul@sjsu.edu}\ \ \ \  
Mark Stamp\thanks{mark.stamp@sjsu.edu}\markonly{\sscoin}
}

\date{}

\begin{document}

\maketitle

\vglue-0.35in

\affil{\sscoin}{Department 
of Computer Science,
San Jose State University,
San Jose, California}

\abstract
Malware detection is a critical aspect of information security. One difficulty that arises 
is that malware often evolves over time. 
To maintain effective malware detection, it is necessary to determine when 
malware evolution has occurred so that appropriate countermeasures can be taken. 
We perform a variety of experiments aimed at detecting points in time where a malware 
family has likely evolved, and we consider secondary tests 
designed to confirm that evolution has actually occurred. 
Several malware families are analyzed, each of which 
includes a number of samples collected over an extended period of time. 
Our experiments indicate that improved results are obtained
using feature engineering based on word
embedding techniques.
All of our experiments are based on machine learning models, 
and hence our evolution detection strategies 
require minimal human intervention and can easily be automated.

\section{Introduction}\label{chap:introduction}

Malware is malicious software that causes disruption in normal activity, 
allows access to unapproved resources, gathers private data of users, 
or performs other improper activity~\cite{Aycock}. 
Developing measures to detect malware 
is a critical aspect of information security.

Malware often evolves due to changing goals of malware developers, 
advances in detection, and so on~\cite{BaratMarius2013Asoc}. 
This evolution can occur through a wide variety of modifications to the code. 
It is essential to detect and analyze malware evolution so that appropriate 
measures can be taken to maintain and improve the effectiveness of detection 
techniques~\cite{BaiJin2014TAoM}.

An obvious technique for analyzing malware evolution consists of 
reverse engineering a large number of samples over an extended period of time,
which is a highly labor-intensive process. Other approaches
to malware evolution include graph 
pruning techniques~\cite{GuptaA2009Aeso}
and analysis of PE file features using support vector 
machines (SVM)~\cite{WadkarMayuri2019MMEU}. 
This latter research shows considerable promise, and 
has the advantage of being fully automated, with no reverse engineering or
other time-consuming analysis required. Our proposed research can be viewed 
as an extension of---and improvement on---the groundbreaking work 
in~\cite{WadkarMayuri2019MMEU}.

We consider several experiments that are designed to detect points in time where 
a malware family has likely evolved significantly. We then perform further experiments 
to confirm that such evolution has actually occurred. All of our experiments have been 
conducted using a significant number of malware families, most of which include 
a large number of samples collected over an extended period of time. 
Furthermore, all or our experiments are based on machine learning,
and hence fully automatable.

For a given malware family, we first separate the available 
samples based on windows of time. We have extracted opcode sequences
from every sample and we use these opcodes as 
features for detecting malware evolution. We experiment with a variety of
feature engineering techniques, and in each case we train linear SVMs
over sliding windows of time. The SVM weights of these models are
compared based on a~$\chi^2$ distance measure, which enables us to
detect changes in the SVM models over time. A point in time where a spike is observed
in the~$\chi^2$ graph shows a substantial
change in SVM models---which indicates a possible evolutionary branch 
in the malware family under consideration. To confirm that such evolution
has actually occurred, we train hidden Markov models (HMM) on either side 
of a significant spike in the~$\chi^2$ graph. 
If a clear distinction between these HMMs is observed, 
it serves as confirmation that significant evolution has been detected.
The primary objective of this research is to implement and analyze 
different variants of this proposed malware evolution detection technique.

The remainder of this paper is organized as follows.
In Section~\ref{chap:background} we discuss relevant related work
in the area of malware evolution. 
Section~\ref{chap:implementation} provides an overview 
of the dataset that we use, as well as brief introductions to the
various machine learning models and techniques that we 
use in this research. We present our experimental results
in Section~\ref{chap:results}. The paper concludes with Section~\ref{chap:conclusion},
where we also outline possible avenues for future work. 

\section{Related Work}\label{chap:background}

Relative to the vast malware research literature, comparatively little 
has been done in the area of malware evolution. In this section, we 
provide a selective review of research related to malware evolution.
 
The malware evolution research in~\cite{GuptaA2009Aeso} is based on large and 
diverse malware dataset that spans nearly two decades. This work focuses on
inheritance properties of malware, and the technique is based on graph pruning. 
The authors claim that many specific traits of various families in their dataset have 
been ``inherited'' from other families. However, it is not entirely clear that 
these ``inherited'' traits are actually inherited, as opposed to having been 
developed independently. In addition, the graph-based analysis 
in~\cite{GuptaA2009Aeso} requires ``extensive manual investigation,'' 
which is in stark contrast to the automated techniques that are considered in this paper. 

The authors of~\cite{MercaldoFrancesco2018Aeso} extract a variety of features 
from Android malware samples and determine trends based on standard software quality
metrics. These results are compared to a similar analysis of trends in 
Android non-malware, or goodware. This work shows that the trends in Android
malware and goodware are fairly similar, indicating that the ``improvement''
in this type of malware has followed a similar path as that of goodware.

The paper~\cite{Ouellette} is focused on detecting new malware variants, 
which is closely related to an evolution problem. The authors 
considered malware variants that would typically defeat machine learning 
based detectors. Their approach relies an an extensive feature set and 
employs semi-supervised learning. In comparison, the approach in this 
paper relies entirely on unsupervised techniques, and we are able to
detect less drastic code modifications.

The work in~\cite{ChenZhongqiang2012Mcat} is nominally focused on malware 
taxonomy. However, this work also provides insight into malware evolution, 
in the form of ``genealogical trajectories.'' The work relies on a variety of
features and uses support vector machines (SVM) for classification. 

We note in passing that machine learning models are trained on features. 
Thus, extracting appropriate features from a dataset is a crucial
step in any malware analysis technique that is based on machine learning. 
We can broadly classify features as static and dynamic---features that can 
be obtained without executing the code are said to be static,
while those that require code execution or emulation are known as dynamic. 
In general, static features are more efficient to collect, 
whereas dynamic features can be more informative and are
typically more robust~\cite{DamodaranAnusha2017Acos}.

The author in~\cite{WadkarMayuri2019MMEU} use static PE file features 
of malware samples as the basis for their malware evolution research. 
Based on these feature, linear SVMs are trained over various time windows 
and the resulting model weights are compared using a~$\chi^2$ distance. 
A spike in the~$\chi^2$ distance graph is shown to be indicative of 
an evolutionary change in a malware family.
Note that this approach is easily automated, 
with no reverse engineering required. 

The research presented in this paper extends and expands on the 
work in~\cite{WadkarMayuri2019MMEU}. 
As in~\cite{WadkarMayuri2019MMEU}, 
we use SVMs together with~$\chi^2$ distance as a means of
detecting evolutionary change. We make several important
contributions that greatly increase the utility of this basic approach.
The novelty of our work includes the use of 
more sensitive static features---we use opcodes as 
compared to derived PE file features---and
we employ various feature engineering techniques. In addition,
we develop an HMM-based secondary test to verify the putative evolutionary changes
obtained from the SVM together with a~$\chi^2$ distance.

\section{Implementation}\label{chap:implementation}

In this section, we give a broad summary of the malware families that comprise 
the dataset used in the research. We also discuss the features and machine 
learning techniques used in our experiments. These features and techniques
forms the basis of our evolutionary experiments in Section~\ref{chap:results}.

\subsection{Dataset}

A malware family represents a collection of samples that have major traits in common. 
Over time, successful malware families will tend to evolve, as malware writers 
develop new features and find different applications for the code base. 

The research in this paper is based of a malware dataset consisting of 
Windows portable executable (PE) files.
From a large dataset, we have extracted~11,037 samples 
belonging to~15 distinct malware families. 
Table~\ref{tab:1} lists these malware families and the number of 
samples per family that we use in our experiments.

\begin{table}[!htb]
    \caption{Number of samples used in experiments}\label{tab:1}
    \vglue 0.1in
    \centering
    \begin{tabular}{c|r|c}
    \midrule\midrule
        \textbf{Family} & \multicolumn{1}{c|}{\textbf{Samples}} & \textbf{Years} \\
        \midrule
        Adload & 791\ \ \  & 2009--2011 \\
        BHO & 1116\ \ \   & 2007--2011\\
        Bifrose & 577\ \ \  & 2009--2011\\
        CeeInject & 742\ \ \  & 2009--2012\\
        DelfInject & 401\ \ \  & 2009--2012\\
        Dorkbot & 222\ \ \  & 2005--2012\\
        Hupigon & 449\ \ \  & 2009--2011\\
        IRCBot & 59\ \ \  & 2009--2012\\
        Obfuscator & 670\ \ \  & 2004--2017\\
        Rbot & 127\ \ \  & 2001--2012\\
        VBInject & 2331\ \ \  & 2009--2018\\
        Vobfus & 700\ \ \  & 2009--2011\\
        Winwebsec & 1511\ \ \  & 2008--2012\\
        Zbot & 835\ \ \  & 2009--2012\\
        Zegost & 506\ \ \  & 2008--2011\\ \midrule
        Total & 11,037\ \ \  & --- \\
        \midrule\midrule
    \end{tabular}
\end{table}

Our Winwebsec and Zbot malware samples were acquired from the 
Malicia dataset~\cite{NappaAntonio2015TMdi}, while the remaining thirteen 
families were extracted from a vast malware dataset that was collected
as part of the work reported in~\cite{KimSamuel2018PHAf}. 
This latter dataset is greater than half a terabyte in size and contains 
on the order of~500,000 malware executables. Our datasets is
available from the authors upon request. 

Most of the malware families that were chosen for this research have a
substantial number of samples available over an extended time period.
The smaller families (e.g., IRCBot and Rbot) were chosen to test the 
our analysis techniques in cases where the training data is severely limited.

As a pre-processing step, we have organized all the malware samples in 
each family according to their creation date. During this initial data wrangling phase, 
any sample having an altered compilation or creation date was 
discarded. 

Next, we briefly discuss each of the malware families in our dataset. 
Note that these families represent a wide variety of types of malware, 
including Trojan, worm, adware, backdoor, and so on.

\begin{description}

\item[\bf Bifrose]\hspace*{-12pt} is a backdoor Trojan that allows an attacker to connect to a remote IP 
using a random port number. Some variants of Bifrose have the capability to hide 
files and processes from the user. Bifrose enables an attacker to view system information, 
retrieve passwords, or execute files by gaining remote control of 
an infected system~\cite{Bifrose}.

\item[\bf CeeInject]\hspace*{-12pt} serves to shielding nefarious activity from detection. 
For example, CeeInject can obfuscate a bitcoin mining client, 
which might be installed on a system to mine bitcoins without the user’s 
knowledge~\cite{CeeInject}.

\item[\bf DelfInject]\hspace*{-12pt} is a worm that enters a system from a file passed by other malware, 
or as a file downloaded accidentally by a client when visiting malignant sites. 
DelfInject drops itself onto the system using an arbitrary document name 
(e.g., \texttt{xpdsae.exe}) and alters the relevant registry entry so that it runs 
at each system start. The malware then injects code into \texttt{svchost.exe} 
so that it can create a connection with specific servers and download files~\cite{DelfInject}.

\item[\bf Dorkbot]\hspace*{-12pt} is a worm that steals user names and passwords by tracking
online activities. It blocks security update websites and can launch denial of service (DoS) 
attacks. Dorkbot is spread via instant messaging applications, social networks, 
and flash drives~\cite{Dorkbot}.

\item[\bf Hotbar]\hspace*{-12pt} is an adware program that may be unintentionally downloaded by a user 
from a malicious website. Being adware, Hotbar displays advertisements 
as the user browses the web~\cite{Hotbar}.

\item[\bf Hupigon]\hspace*{-12pt} is a family of backdoor Trojans. This malware opens a backdoor 
server enabling other remote computers to control a compromised system~\cite{Hupigon}.

\item[\bf Obfuscator]\hspace*{-12pt} hides its purpose through obfuscation. The underlying malware can 
have virtually any payload~\cite{Obfuscator}.

\item[\bf Rbot]\hspace*{-12pt} is a backdoor Trojan that enables an attacker to control an infected 
computer using an IRC channel. It then spreads to other computers by scanning 
for network shares and exploiting vulnerabilities in the system. Rbot includes many 
advanced features and it has been used to launch DoS attacks~\cite{Rbot}.

\item[\bf VBInject]\hspace*{-12pt} primarily serves to disguise other malware. VBInject is a 
packaged malware, i.e., malware that utilizes techniques of encryption 
and compression to obscure its contents. This makes it difficult to recognize 
malware that it is concealing. VBInject was first seen in~2009 and appeared again 
in~2010~\cite{VBinject}.

\item[\bf Vobfus]\hspace*{-12pt} is a malware family that downloads other malware onto a user’s computer. 
It uses the Windows \texttt{autorun} feature to spread to other devices such as flash drives. 
Vobfus makes long lasting changes to the device configuration that cannot be restored 
simply by removing the malware from the system~\cite{Vobfus}.

\item[\bf Winwebsec]\hspace*{-12pt} is a Trojan that presents itself as an antivirus software. It shows 
misleading messages to the users stating that the device has been infected and attempts 
to persuade the user to pay to remove these non-existent threats~\cite{Winwebsec}.

\item[\bf Zbot]\hspace*{-12pt} is a Trojan that attempts to steal confidential information from a 
compromised computer. It explicitly targets system data, online sensitive data, 
and banking information, and it can also be easily modified to accumulate 
other kinds of data. The Trojan itself is generally disseminated through drive-by 
downloads and spam campaigns. Zbot was originally discovered in 
January~2010~\cite{Zbot}.

\item[\bf Zegost]\hspace*{-12pt} is a backdoor Trojan that injects itself into \texttt{svchost.exe}, 
thus allowing an attacker to execute files on the compromised system~\cite{Zegost}.

\end{description}

\subsection{Feature Extraction}

Mnemonic opcodes are machine-level language instructions that specify a particular 
operation that is to be performed~\cite{RezaeiSaeid2016Mduo}. For this research, 
our dataset consists of malware samples in the form of portable executable (PE) files.
The primary feature that we consider are opcode sequences extracted 
from these executable files. 
We have also segregated the malware samples in each family according 
to their creation date. 

\subsection{Classification Techniques}

In this section we will provide an overview of each machine learning
technique that we employ in this research. Additional pointers 
to the literature are provided.

\subsubsection{Support Vector Machines}

Support vector machines (SVM) are one of the most popular classes of machine
learning techniques. An SVM attempts to find a separating hyper-plane between two 
labeled classes of data~\cite{StampMark2018Itml}. By utilizing the so-called 
``kernel trick'', an SVM can map the input data to a high-dimensional space 
where the additional space can afford a greater opportunity to find a separating 
hyperplane. The ``trick'' of the kernel trick is that this mapping does not result in
any significant increase in the work factor. 

A linear SVM assigns a well-defined weight to each feature in the training vector. 
These weights specify the relative importance that the SVM places on each feature 
which can serve as a useful when ranking the importance of features.
In our experiments, we rely heavily on this aspect of linear SVMs.

Analogos to~\cite{WadkarMayuri2019MMEU},
in our experiments, we define the two classes of an SVM as follows.
All the  samples within the most recent one-year time window 
are class ``$+1$'', while all samples from the current month are defined as 
class ``$-1$''. For example, in Table~\ref{tab:2} we give three consecutive time 
windows, along with the time frames corresponding to the two classes in each case.

\begin{table}[!htb]
    \caption{Sliding time window example}\label{tab:2}
    \vglue 0.1in
    \centering
    \begin{tabular}{c|c|c}
    \midrule\midrule
        \textbf{Time Window} & \textbf{Class} $+1$ & \textbf{Class} $-1$\\
        \midrule
        Jan. 2011--Jan. 2012 & Jan. 2011--Dec. 2011 & Jan. 2012 \\
        Feb. 2011--Feb. 2012 & Feb. 2011--Jan. 2012 & Feb. 2012 \\
        Mar. 2011--Mar. 2012 & Mar. 2011--Feb. 2012 & Mar. 2012\\
        \midrule
    \end{tabular}
\end{table}

\subsubsection{$\chi^2$ Statistic}

The~$\chi^2$ statistic is a normalized sum of square deviation between the 
observed and expected frequency distributions. This statistic is calculated as
\[
  {\chi}^2=\sum_{i=1}^{n} \frac{(o_i - e_i)^2}{e_i}
\]
where~$n$ denotes the number of features or observations, 
$o_i$ is the observed value of the $i^{\thth}$ instance, 
and~$e_i$ is expected value of the $i^{\thth}$ instance.

For our experiments, this statistic is used to quantify the differences between 
SVM feature weights of different models, where these models were trained over
different time windows. As mentioned above, 
we use a time period of one year for one class, 
and a time window of the following month as the other class. 
We compute this~$\chi^2$ ``distance'' between pairs of models that 
are trained on overlapping time windows. Any points in the resulting~$\chi^2$ graph 
where a substantial change (i.e., a ``spike'') occurs indicates a
point where adjacent SVM models differ significantly. 
These are points of interest, since they indicate the times at which the 
code has likely been substantially modified.

\subsubsection{Word2Vec}

Word2Vec is a ``word'' embedding technique
that can be applied more generally to features. Word2Vec 
is extremely popular in language modeling. The embedding vectors
produced by state-of-the-art implementations of Word2Vec
capture a surprising level of the semantics of a language. 
That is, words that are similar in meaning
are ``close'' in in the Word2Vec embedding space~\cite{PopovIgor2017Mdum}. 
An oft-cited example of the strength of Word2Vec is the following.
If we let
$$
  w_0=\mbox{``king''}, w_1=\mbox{``man''}, w_2=\mbox{``woman''}, w_3=\mbox{``queen''}
$$
and~$V(w_i)$ is the Word2Vec embedding of the word~$w_i$, then~$V(w_3)$
is the vector that is closest---in terms of cosine similarity---to
$$
  V(w_0) - V(w_1) + V(w_2)
$$

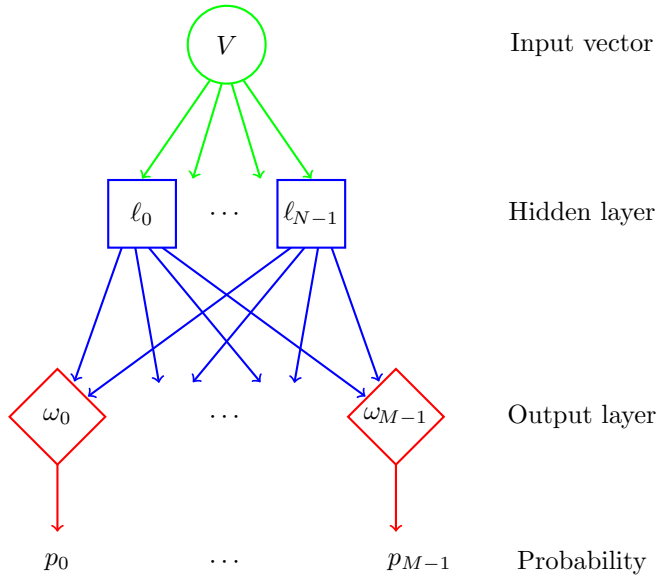
\begin{figure}[!htb]
  \centering
\begin{tikzpicture}[scale=0.9]
    
    \draw[thick,color=green] (4.25,8.5) circle (0.575);

    \draw[thick,color=blue] (2.5,5.5) rectangle (3.5,6.5);
    \node at (4.25,6.0) {$\cdots$};
    \draw[thick,color=blue] (5.0,5.5) rectangle (6.0,6.5);
        
    \draw[thick,color=red,rotate around={45:(1.75,3.0)}] (1.25,2.5) rectangle (2.25,3.5);
    \draw[thick,color=red,rotate around={45:(6.75,3.0)}] (6.25,2.5) rectangle (7.25,3.5);

    \draw[thick,color=green,->] (4.0,7.97) -- (3.0,6.52);
    \draw[thick,color=green,->] (4.175,7.92) -- (3.75,6.52);
    \draw[thick,color=green,->] (4.325,7.92) -- (4.75,6.52);
    \draw[thick,color=green,->] (4.5,7.97) -- (5.5,6.52);
    
    \draw[thick,color=blue,->] (2.7,5.5) -- (2.0,3.52);
    \draw[thick,color=blue,->] (2.9,5.5) -- (3.25,3.5);
    \draw[thick,color=blue,->] (3.1,5.5) -- (4.75,3.51);
    \draw[thick,color=blue,->] (3.3,5.5) -- (6.3,3.3);

    \draw[thick,color=blue,->] (5.2,5.5) -- (2.2,3.3);
    \draw[thick,color=blue,->] (5.4,5.5) -- (3.75,3.51);
    \draw[thick,color=blue,->] (5.6,5.5) -- (5.25,3.5);
    \draw[thick,color=blue,->] (5.8,5.5) -- (6.5,3.52);

    \draw[thick,color=red,->] (1.75,2.3) -- (1.75,1.3);
    \draw[thick,color=red,->] (6.75,2.3) -- (6.75,1.3);

    \node at (4.25,8.5) {$V$};

    \node at (3.0,6.0) {$\ell_0$};
    \node at (5.5,6.0) {$\ell_{\kern-1pt N-1}$};

    \node at (1.75,3.0) {$\omega_0$};
    \node at (4.25,3.0) {$\cdots$};
    \node at (6.75,3.0) {$\omega_{\kern-1pt M-1}$};

    \node at (1.75,0.85) {$p_0$};
    \node at (4.25,0.85) {$\cdots$};
    \node at (7.1,0.85) {$p_{M-1}$};
    
    \node at (9.5,8.5) {Input vector};
    \node at (9.5,6.0) {Hidden layer};
    \node at (9.5,3.0) {Output layer};
    \node at (9.5,0.85) {Probability};

\end{tikzpicture}
  \caption{Neural network to obtain Word2Vec embeddings}\label{fig:w2v}
\end{figure}

Word2Vec is based on a shallow, two-layer neural networks,
as illustrated in Figure~\ref{fig:w2v}. Training such a model consists of
determining the weights~$w_i$ based on a large training corpus~\cite{Word2Vec}.
These weights yield the Word2Vec embedding vectors.

In this paper, we compute Word2Vec embeddings 
based on extracted opcode sequence from malware samples. 
These word embeddings are then used as features in SVM classifiers.
In this context, we can view the use of Word2Vec as a form of 
feature engineering.

One of the great strengths of Word2Vec is that training is extremely efficient.
The key tricks that enable efficient training of such models are 
subsampling of frequent words, and so-called negative sampling,
whereby only a subset of the weights that are affected by
a training pair are adjusted at each iteration. For additional
details on Word2Vec, see, for example~\cite{Stamp19alpha}.

\subsubsection{Hidden Markov Models} 

A Markov process is a statistical model that has states with known and fixed
probabilities of state transitions. A hidden Markov model (HMM) extends this
concept to the case where the states are ``hidden,'' in the sense that
they are not directly observable. 

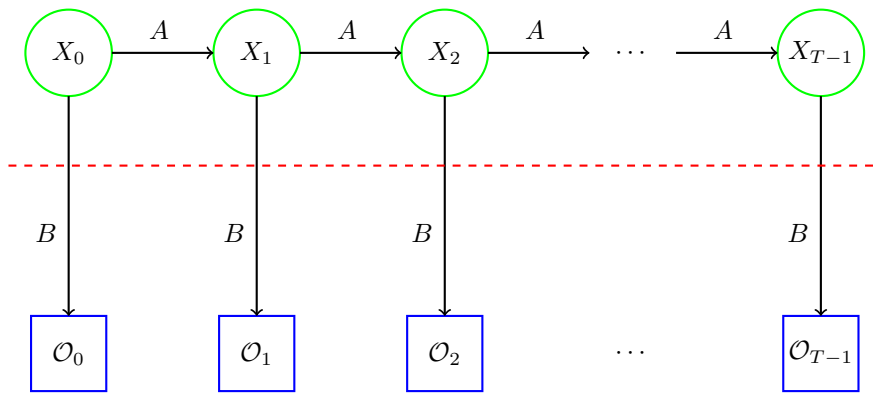
\begin{figure}[!htb]
    \centering
    \begin{tikzpicture}[scale=1.0]
    
    \draw[thick,color=blue] (0,0) rectangle (1,1);
    \draw[thick,color=blue] (2.5,0) rectangle (3.5,1);
    \draw[thick,color=blue] (5,0) rectangle (6,1);
    \draw[thick,color=blue] (10,0) rectangle (11,1);

    \draw[thick,color=green] (0.5,4.5) circle (0.575);
    \draw[thick,color=green] (3,4.5) circle (0.575);
    \draw[thick,color=green] (5.5,4.5) circle (0.575);
    \draw[thick,color=green] (10.5,4.5) circle (0.575);
    
    \node at (0.5,0.5){$\O_0$};
    \node at (3,0.5){$\O_1$};
    \node at (5.5,0.5){$\O_2$};
    \node at (8,0.5){$\cdots$};
    \node at (10.5,0.5){$\O_{T-1}$};

    \node at (0.5,4.5){$X_0$};
    \node at (3,4.5){$X_1$};
    \node at (5.5,4.5){$X_2$};
    \node at (8,4.5){$\cdots$};
    \node at (10.5,4.5){$X_{T-1}$};
       
    \node at (1.7,4.8){$A$};
    \node at (4.2,4.8){$A$};
    \node at (6.7,4.8){$A$};
    \node at (9.2,4.8){$A$};
    
    \node at (0.2,2.1){$B$};
    \node at (2.7,2.1){$B$};
    \node at (5.2,2.1){$B$};
    \node at (10.2,2.1){$B$};
    
     \draw[thick,color=black,->] (1.075,4.5) -- (2.425,4.5);
     \draw[thick,color=black,->] (3.575,4.5) -- (4.925,4.5);
     \draw[thick,color=black,->] (6.075,4.5) -- (7.425,4.5);
     \draw[thick,color=black,->] (8.575,4.5) -- (9.925,4.5);

     \draw[thick,color=black,->] (0.5,3.925) -- (0.5,1);
     \draw[thick,color=black,->] (3.0,3.925) -- (3.0,1);
     \draw[thick,color=black,->] (5.5,3.925) -- (5.5,1);
     \draw[thick,color=black,->] (10.5,3.925) -- (10.5,1);

    \draw[thick,dashed,color=red] (-0.3,3) -- (11.2,3);
   
\end{tikzpicture}
    \caption{Hidden Markov Model}\label{fig:6}
\end{figure}

Figure~\ref{fig:6} provides a generic view of an HMM. Here, the states~$X_i$
are determined by the row stochastic~$N\times N$ matrix~$A$. The states~$X_i$
are not directly observable, but as the name implies, the observations~$\O_i$ 
can be observed. The hidden states are probabilistically related to the observations 
via the~$N\times M$ row stochastic matrix~$B$. Here, $N$ is the number of
hidden states of the model, and~$M$ is the number of distinct observation symbols,
and~$T$ in Figure~\ref{fig:6} is the length of the observation sequence.
There is also a row stochastic initial state distribution matrix, which is
denoted as~$\pi$. The three matrices, $A$, $B$, and~$\pi$ define an HMM,
and we adopt the notation~$\lambda = (A, B, \pi)$.

The following three HMM problems can be solved efficiently:

\begin{description}
\item[\bf Problem 1]
Given a model $\lambda = (A, B, \pi)$ and an observation sequence~$\O$, we
need to find~$P(\O\,|\,\lambda)$. That is, an observation sequence can be 
scored to see how well it fits a given model.

\item[\bf Problem 2]
Given a model $\lambda = (A, B, \pi)$ and an observation sequence~$\O$, we
can determine an ``optimal'' hidden state sequence. In the HMM sense,
optimal means that we maximize the expected number of correct states.
This is an expectation maximization (EM) algorithm.

\item[\bf Problem 3]
Given~$\O$ and a specified~$N$, we can determine a model~$\lambda = (A, B, \pi)$ 
that maximizes~$P(\O\,|\,\lambda)$. This is, we can training of a model to fit a
given observation sequence.
\end{description}

In this research, we employ the algorithms for problems~1 and~3 above. 
That is, we train HMMs, and we use trained HMMs to score samples.

\subsubsection{Experimental Approach}\label{sect:expApp}

To automatically determine points in time where significant evolutionary changes occur 
in malware families, we tag each sample in the family according to the date on which it 
was compiled. We also extract the opcode sequences from each of the malware samples. 

As a first set of experiments, we train a series of linear SVMs directly on the 
extracted opcodes, as discussed above. We
then attempt to improve on these results by
considering several feature engineering techniques,
in all cases using linear SVM weights and~$\chi^2$ graphs.

For our first attempt at feature engineering, we consider opcode $n$-grams.
We then experiments using Word2Vec embeddings of the opocdes. 
Finally, we repeat the word embedding experiments, but
based on the~$B$ matrices obtained from trained HMMs,
instead of Word2Vec embeddings. 
We refer to this HMM-based word embedding technique as HMM2Vec.

As discussed above, when training, one class consists of all samples belonging 
to a specific family within a one-year time window, while the other class consists 
of the samples from the subsequent one-month time window. Such a model 
contrasts the family characteristics over a one month period to the characteristics 
of the previous one-year time interval. From each such model, we obtain a 
vector of linear SVM weights. Then we shift our time window one month ahead, 
and again train an SVM and obtain another vector of SVM weights.

For each set of experiments, 
we obtain a series of snapshots of the 
samples---in the form of linear SVM weights---based 
on overlapping sliding windows, where each SVM is
trained over a one year time-frame. Adjacent SVM weight vectors
are based on one-month offsets. We use these SVM weight vectors 
as a basis for tracking changes in the underlying models,
and we quantify those changes using the~$\chi^2$ statistics, 
as discussed above. Potential evolutionary points appear as spikes
in the resulting~$\chi^2$ graph.

As a secondary test, for each significant spike in the~$\chi^2$ graph, 
we train two HMMs, one on either side of the spike. 
We then score samples on both sides of the spike using both HMMs.
If the sample scores are observably different for each of these HMMs
on each side of the spike, this serves to confirm that significant
evolutionary change in the malware family has occurred. 

For this secondary test, we can quantify the evolutionary effect by 
computing the~$\chi^2$-like evolution score
$$
    E = \frac{1}{n} \sum_{i=1}^n \frac{\bigl(\widehat{S}(x_i)-S(x_i)\bigr)^2}{S(x_i)}
$$
where $S(x_i)$ is the HMM score of the sample~$x_i$ 
using the ``correct'' model and~$\widehat{S}(x_i)$ is 
the score of~$x_i$ using the ``incorrect'' model.
For example, if sample~$x_i$ occurs before the spike, 
then~$S(x_i)$ is the score obtained using the model
that was trained on data before the spike,
and~$\widehat{S}(x_i)$ is the score of~$x_i$
using the model trained after the spike. The larger
the evolution score~$E$, the stronger the evidence of evolution.
Note that the~$1/n$ factor is needed since the number of samples 
available differs for different families, and the number of samples might
also differ for different spike computations within the same family.

\section{Experiments and Results}\label{chap:results}

In this section, we present and discuss the results of the experiments 
outlined in the previous section.
First, we provide a graphical illustration of 
our HMM-based secondary test. Then we consider
opcode-SVM experiments, followed by analogous SVM
experiments based on opcode $n$-gram features. 
Neither of these techniques produce particularly strong results, and hence
we then turn our attention to additional opcode-based feature engineering.
Specifically, we aply word embedding techniques to opcode sequence
and train SVM classifiers based on these engineered features.
These experiments prove to be highly successful.

\subsection{HMM-Based Secondary Test}

The previous work in~\cite{WadkarMayuri2019MMEU} 
is based on PE file features and uses linear SVM analysis 
to detect evolutionary changes in a malware family. 
We perform similar analysis in this paper, but based on opcode
features, and using word embedding techniques for feature engineering.
In this paper, we also employ hidden Markov models as
a secondary test to confirm suspected evolutionary changes. 

As discussed above, once distinct spikes have been obtained from 
the~$\chi^2$ similarity graph,  we train an HMM model on both sides 
such spikes using extracted opcode sequences. 
Both models are then used for scoring malware samples
on either side of the spike. Here, we illustrate this secondary
test for a specific case.

\begin{figure}[!htb]
	\centering
	\begin{tabular}{ccc}
		\includegraphics[width=0.45\textwidth]{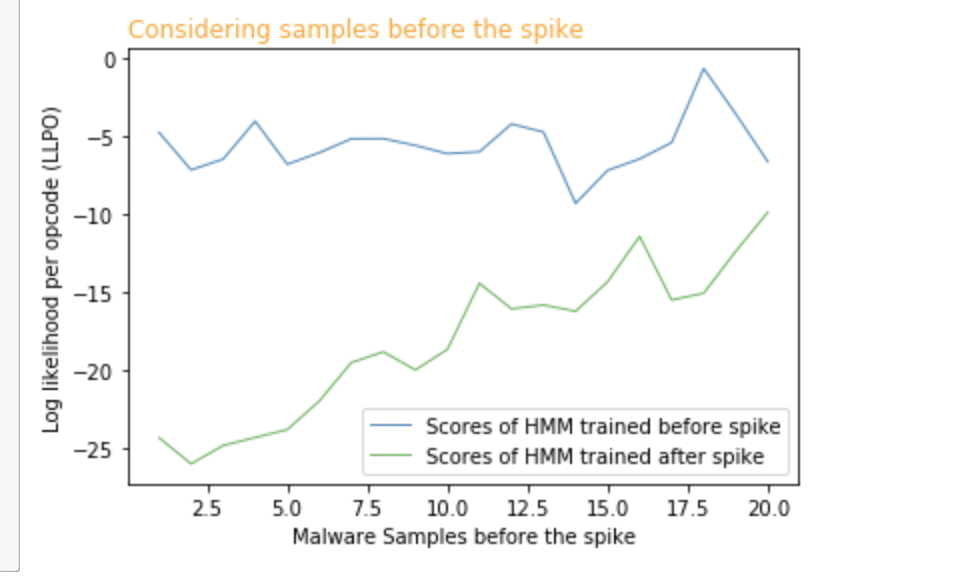}
		&
		\includegraphics[width=0.465\textwidth]{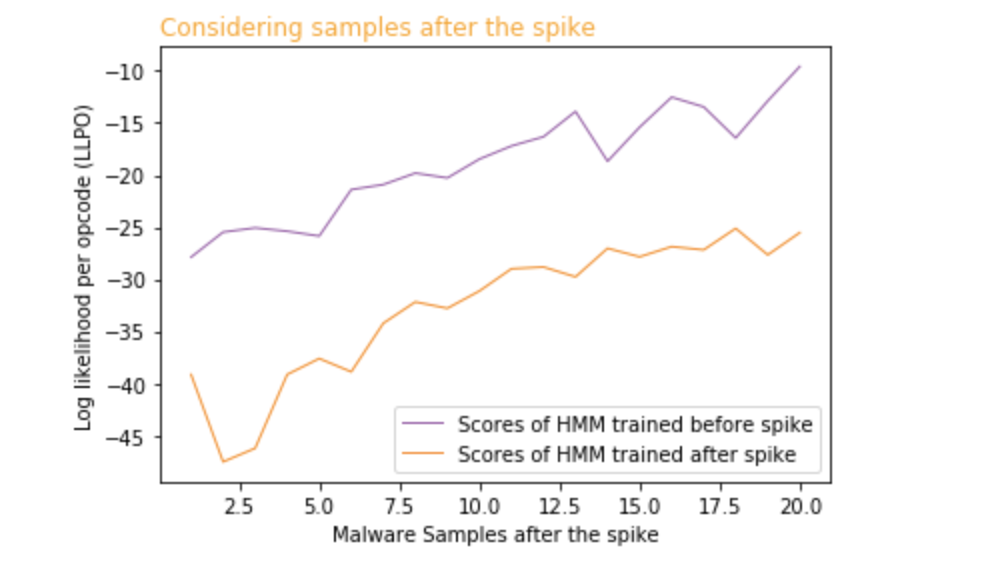}
		\\
		(a) Malware samples before spike
		&
		(b) Malware samples after spike
	\end{tabular}
	\caption{Hidden Markov model trained before and after the spike}\label{fig:7}
\end{figure}

Figure~\ref{fig:7}~(a) depicts the scores obtained when scoring 
samples before a~$\chi^2$ graph spike using both HMMs, 
while Figure~\ref{fig:7}~(b)
gives the corresponding result for samples after the spike.
In all cases, the scores are log likelihood per opcode (LLPO), that is, 
the scores have been normalized so as to be independent of the 
length of the opcode sequence.


In both Figure~\ref{fig:7}~(a) and~(b) we see that the scores are 
distinct for the two models on each sample tested.
These results demonstrates that the model trained before and 
the model trained after the spike are
significantly different which, in turn, indicates that the samples used to train 
the models differ significantly. This is a clear sign of an evolutionary 
branch point in the malware family.

\subsection{Opcode-SVM Results}

In this section, we discuss experiments on~15 malware families,
based on opcode sequences and SVMs. That is, opcodes sequences
are used directly as features in linear SVM models, with the 
resulting model weights used to compute~$\chi^2$ graphs.

\begin{figure}[!htb]
\centering
\includegraphics[width=0.7\textwidth]{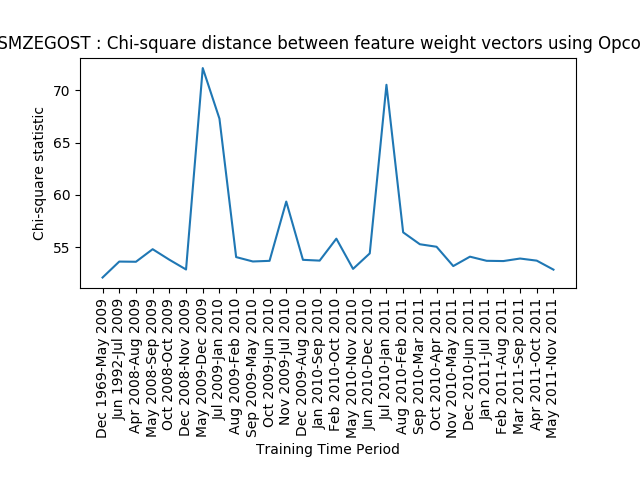}
\caption{Opcode-SVM $\chi^2$ similarity graph for Zegost}
\label{fig:8}
\end{figure}

Figure~\ref{fig:8} gives such a~$\chi^2$ graph for the 
Zegost family, which has~506 malware samples
from the years~2008 through~2011. 
We observe multiple spikes in 
the graph but the secondary HMM test
does not yield impressive results for any of these spikes. 
Hence, we conclude that this particular test
does not reveal any strong evolution result for the Zegost family.        

\begin{figure}[!htb]
   \centering
   \begin{tabular}{ccc}
    \includegraphics[width=0.33\textwidth]{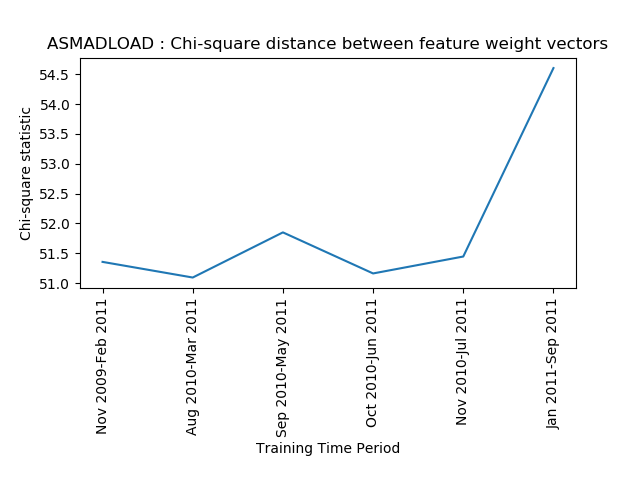}
    &
    \includegraphics[width=0.33\textwidth]{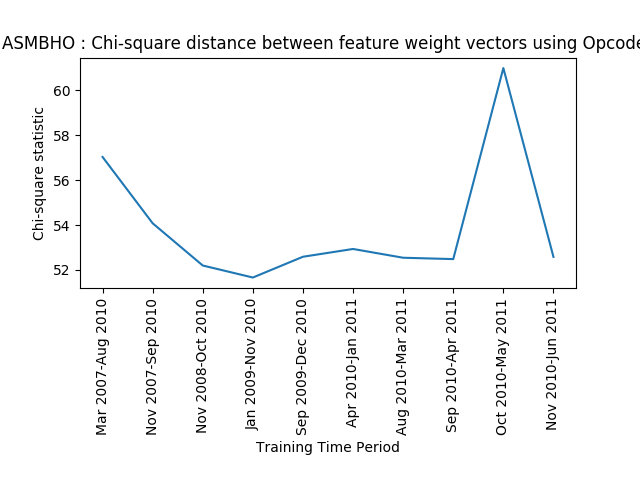}
    &
    \includegraphics[width=0.33\textwidth]{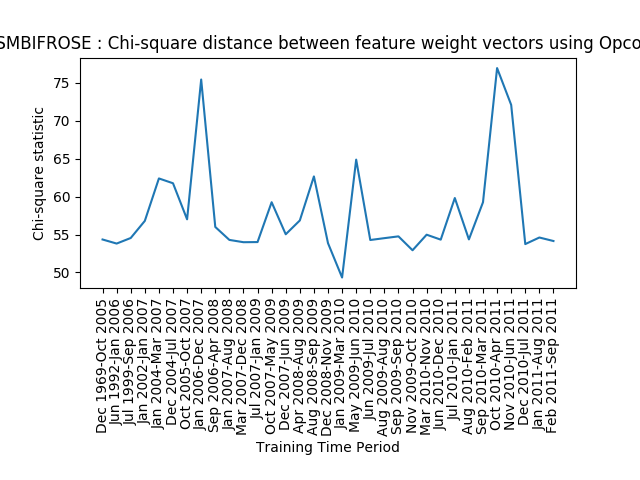}
    \\
    (a) Adload
    &
    (b) BHO
    & 
    (c) Bifrose
    \\
    \includegraphics[width=0.33\textwidth]{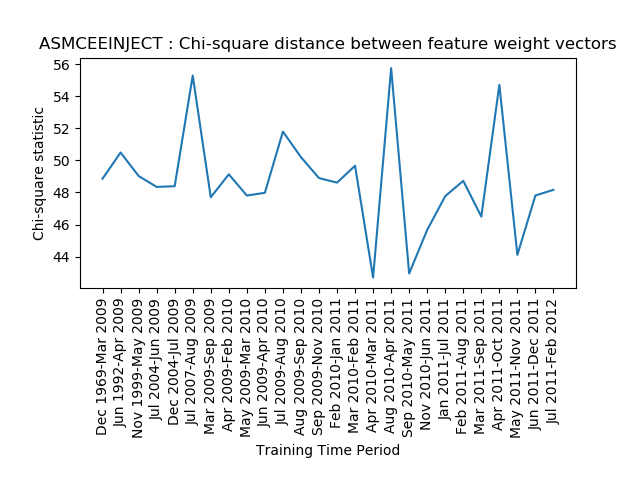}
    & 
    \includegraphics[width=0.33\textwidth]{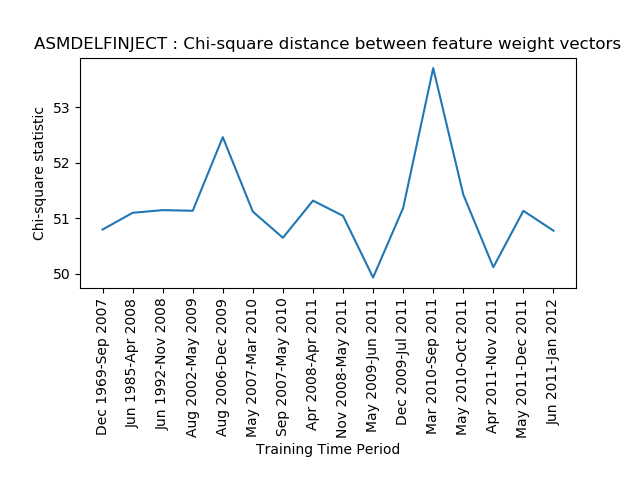}
    &
    \includegraphics[width=0.33\textwidth]{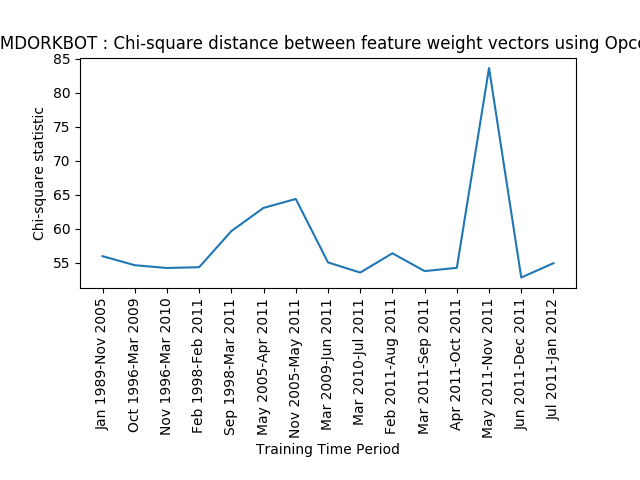}
    \\
    (d) CeeInject
    &
    (e) DelfInject
    &  
    (f) Dorkbot
    \\
    \includegraphics[width=0.33\textwidth]{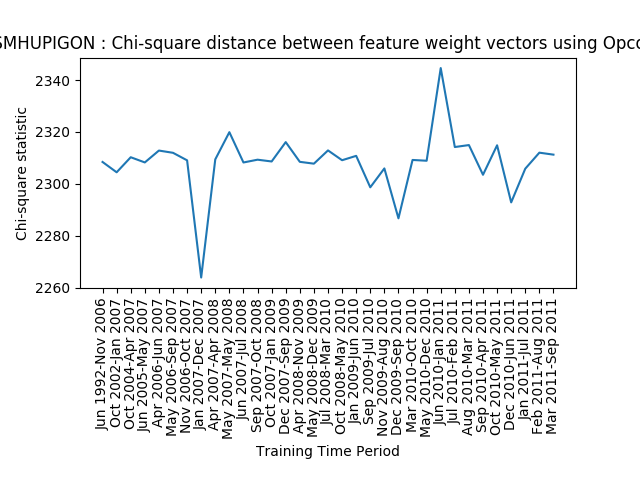}
    &
    \includegraphics[width=0.33\textwidth]{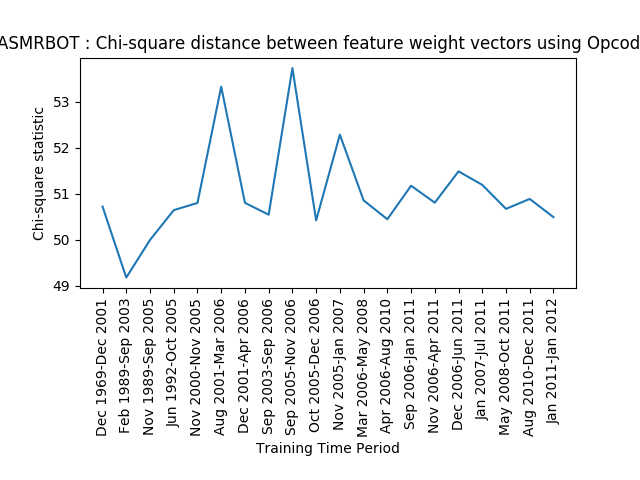}
    &
    \includegraphics[width=0.33\textwidth]{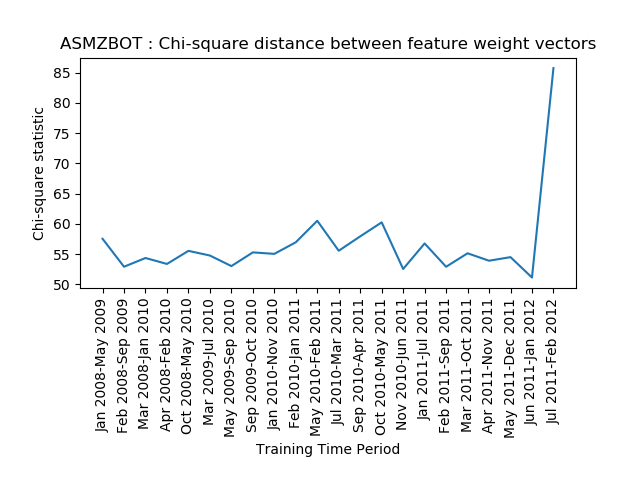}
    \\
    (g) Hupigon
    & 
    (h) Rbot
    &
    (i) Zbot
    \end{tabular}
    \caption{Opcode-SVM $\chi^2$ graphs for selected families}\label{fig:9}
\end{figure}

We also computed opcode-SVM~$\chi^2$ graphs for the 
remaining~14 malware families, nine of which appear in Figure~\ref{fig:9}. 
For Adload and BHO, we observe that the graphs do not show any 
significant spikes except at the last time period. We are not able
to perform the secondary HMM test at the extreme endpoint,
so we are not able to confirm or deny these as evolution points.

From Figure~\ref{fig:9}, we observe considerable fluctuation in the graphs
for Bifrose, CeeInject, Hupigon, and Rbot, but none of these fluctuations stand out as 
clear points of possible evolutionary change in the families. 
That is, for these families, it appears that we only observe background 
noise. 

\subsection{Opcode $n$-gram-SVM Results}

Next, we consider analogous experiments to those of the previous section, but
based on opcode $n$-grams. 
This can be viewed as a first attempt at feature engineering. As with the
previous experiments, we train linear SVMs on these features
and construct~$\chi^2$ graphs.
We consider overlapping~$n$-grams, and we experimented with~$n = 2, 3, 5, 10$
on each of the~15 families.

\begin{figure}[!htb]
   \centering
   \begin{tabular}{ccc}
    \includegraphics[width=0.45\textwidth]{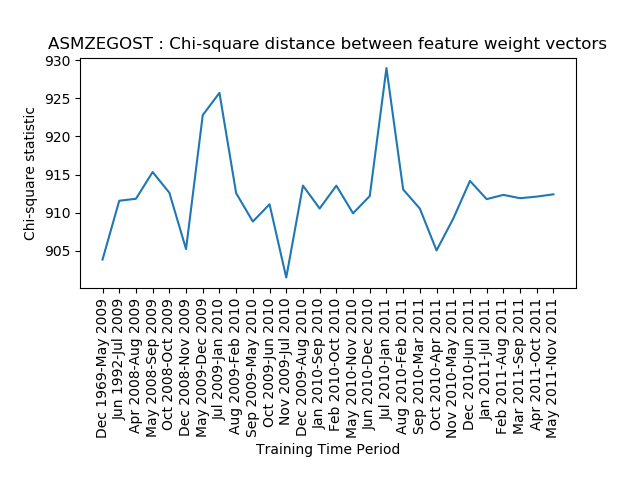}
    &
    &
    \includegraphics[width=0.45\textwidth]{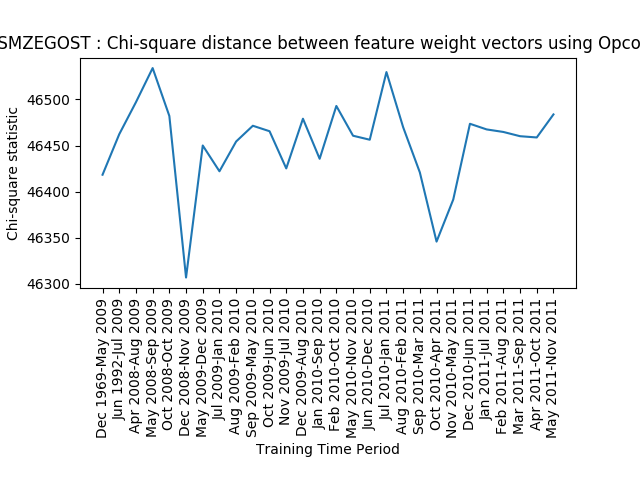}
    \\
    (a) Zegost 2-gram
    &
    &
    (b) Zegost 5-gram
    \end{tabular}
    \caption{Opcode-$n$-gram-SVM~$\chi^2$ graphs for Zegost}\label{fig:10}
\end{figure}

Examples of the results of these~$n$-gram experiments are
given in Figure~\ref{fig:10}, which shows $2$-gram
and~$5$-gram~$\chi^2$ graphs for the Zegost family. 
For both of these cases, we see only noisy results,
with no clear evolutionary points. These results are
typical of our $n$-gram experiments and we
conclude that opcode $n$-grams are not useful for our purposes.

Next, we consider Word2Vec and HMM2Vec word embeddings
These feature engineering techniques
prove to be more effective for detecting evolutionary changes,
with HMM2Vec giving us our strongest results.

\subsection{Opcode-Word2Vec-SVM Results}

Here, we generate Word2Vec models based on opcode sequences. 
We then train linear SVMs over each time window, based on these
Word2Vec embeddings, and we compute~$\chi^2$ graphs of the SVM weights. 
As above, spikes in this graph indicate points in
time where evolution might have occurred.

\begin{figure}[!htb]
   \centering
   \begin{tabular}{ccc}
    \includegraphics[width=0.45\textwidth]{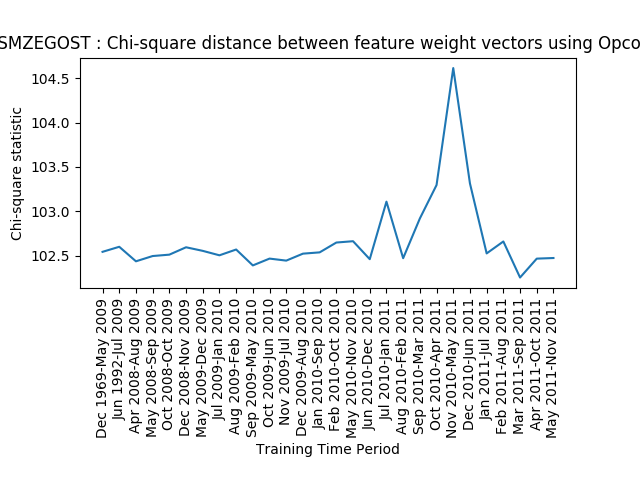}
    &
    &
    \includegraphics[width=0.45\textwidth]{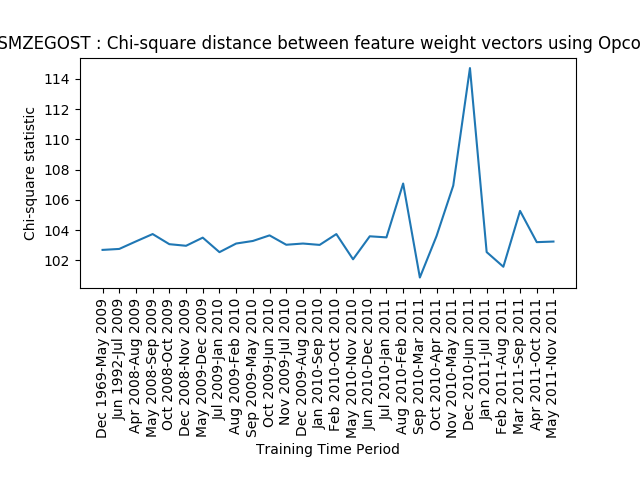}
    \\
    (a) Zegost vector length~2
    &
    &
    (b) Zegost vector length~3
    \end{tabular}
    \caption{Opcode-Word2Vec-SVM~$\chi^2$ graphs for Zegost}\label{fig:12}
\end{figure}

Figures~\ref{fig:12}~(a) and~(b) gives the~$\chi^2$ graphs for the Zegost family 
with Word2Vec embeddings of length~2 and~3,
respectively. From Figure~\ref{fig:12}, we observe that feature
weights in certain time windows diverge significantly from their average values.
Specifically, these time periods are November 2010 and May 2011,
and these are the points in time where significant
evolution in the family may have taken place.
We also note the similarity between the results for embedding
vector lengths~2 and~3. This can be viewed as a sign of the
stability of the underlying approach, and serves to 
provide additional confidence in the putative evolution points.

\begin{figure}[!htb]
    \centering
    \begin{tabular}{ccc}
    \includegraphics[width=0.45\textwidth]{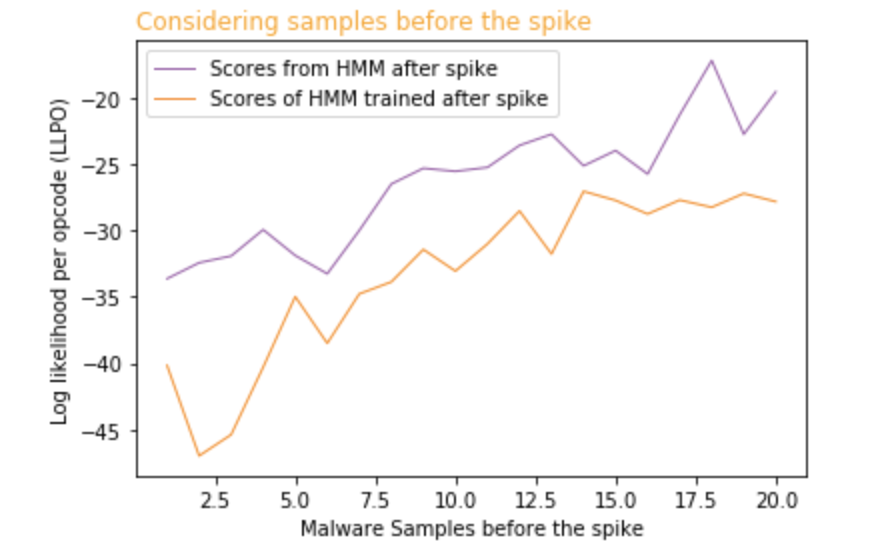}
    & &
    \includegraphics[width=0.45\textwidth]{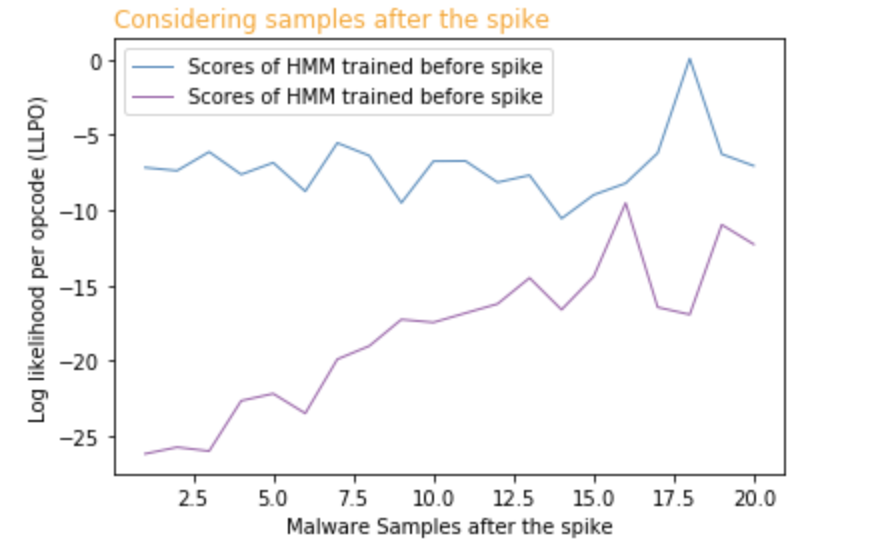}
    \\
    (a) Before spike
    & &
    (b) After spike
    \end{tabular}
    \caption{Zegost HMM secondary test for opcode-Word2Vec-SVM}\label{fig:16}
\end{figure}

Applying our secondary HMM verification technique to the spike in Figure~\ref{fig:12},
we obtain the results in Figure~\ref{fig:16}, which
confirm that the malware family has evolved at this point. 
Since vector lengths of~2 and~3 give us consistent results 
for Zegost, for our remaining Word2Vec experiments, 
we use embedding vectors of length~2 in all cases.

\begin{figure}[!htb]
   \centering
   \begin{tabular}{ccc}
    \includegraphics[width=0.33\textwidth]{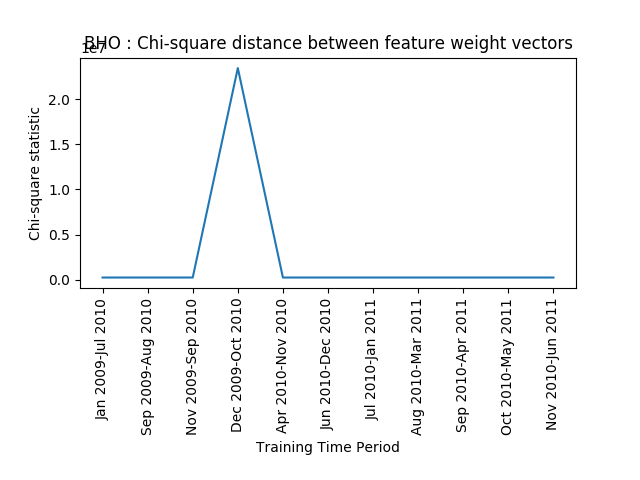}
    & 
    \includegraphics[width=0.33\textwidth]{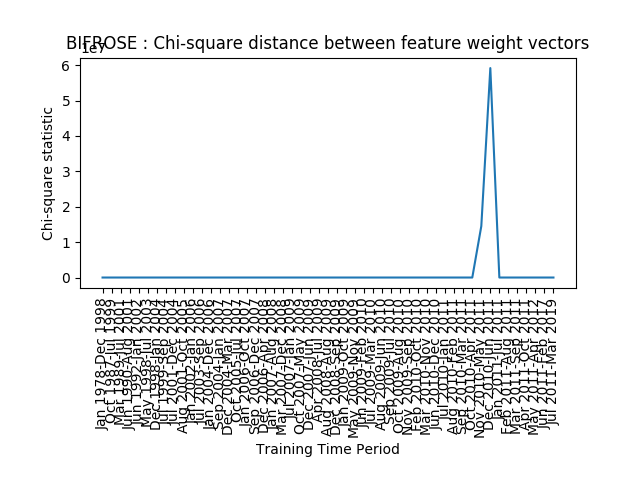}
    &
    \includegraphics[width=0.33\textwidth]{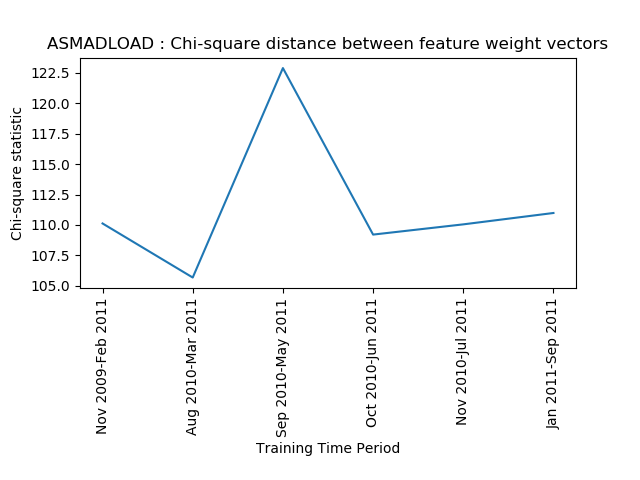}
    \\
    (a) BHO vector size 2
    &
    (b) Bifrose vector size 2
    & 
    (c) Adload vector size 2
    \\
    \includegraphics[width=0.33\textwidth]{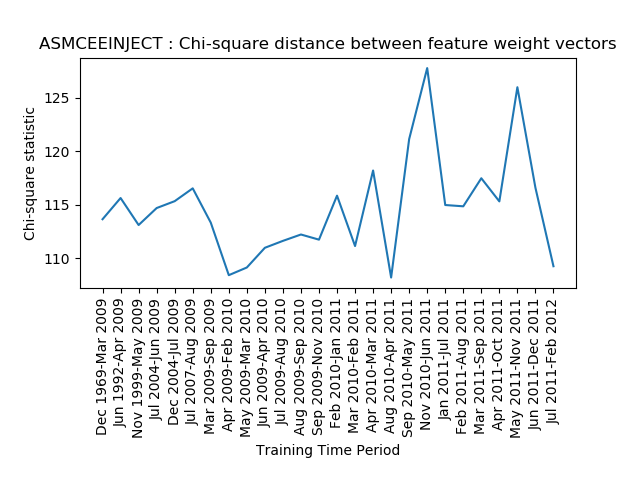}
    & 
    \includegraphics[width=0.33\textwidth]{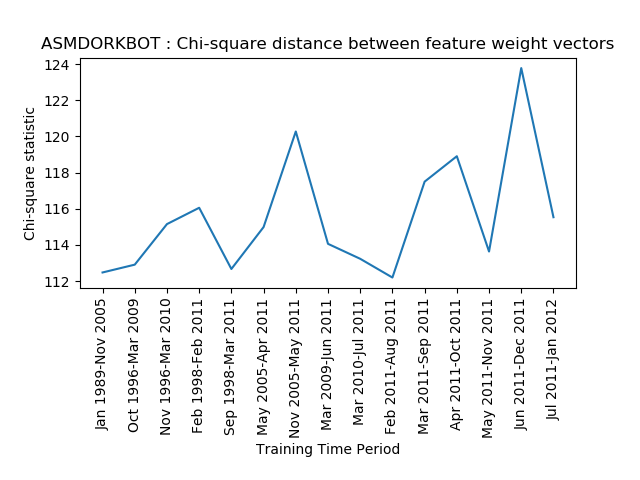}
    &
    \includegraphics[width=0.33\textwidth]{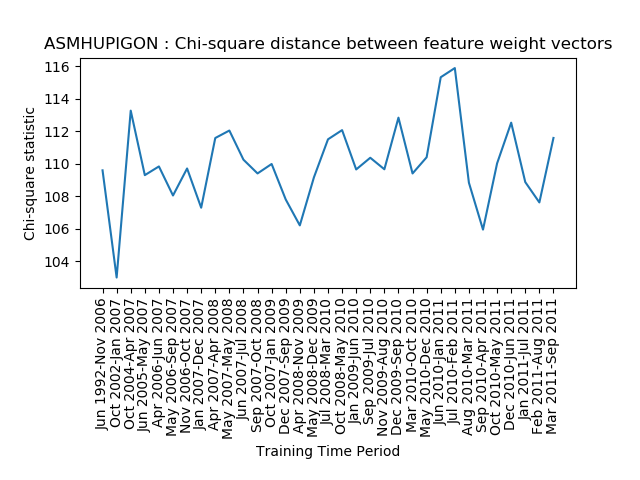}
    \\
    (d) CeeInject vector size 2
    &
    (e) Dorkbot vector size 2
    &  
    (f) Hupigon vector size 2
    \\[3ex]
    \multicolumn{3}{c}{
    \begin{tabular}{cc}
    \includegraphics[width=0.33\textwidth]{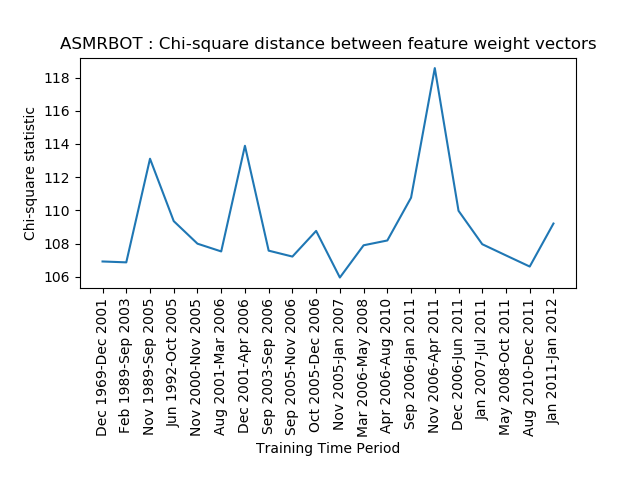}
    &
    \includegraphics[width=0.33\textwidth]{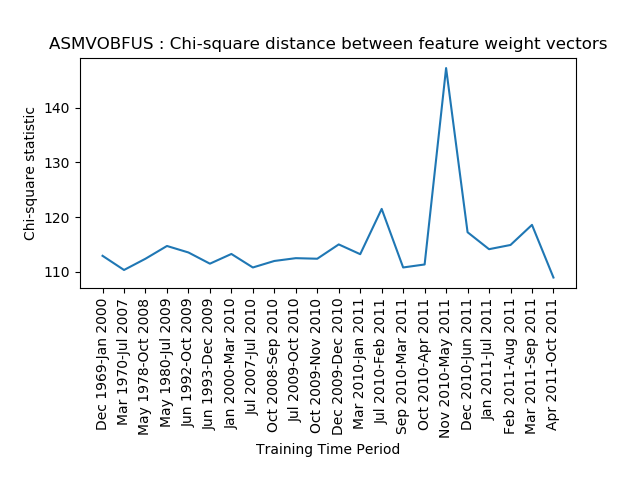}
    \\
    (g) Rbot vector size 2
    & 
    (h) Vobfus vector size 3
    \end{tabular}}
    \end{tabular}
    \caption{Opcode-Word2Vec-SVM~$\chi^2$ graphs for selected families}\label{fig:13}
\end{figure}

Figure~\ref{fig:13} shows our Word2Vec based~$\chi^2$ graphs for eight additional
malware families. Four of these families---BHO, Bifrose, Adload, 
and Vobfus---perform well with this approach, in the sense that we
detect clear spikes in their graphs.

\begin{figure}[!htb]
	\centering
	\begin{tabular}{ccc}
	\includegraphics[width=0.35\textwidth]{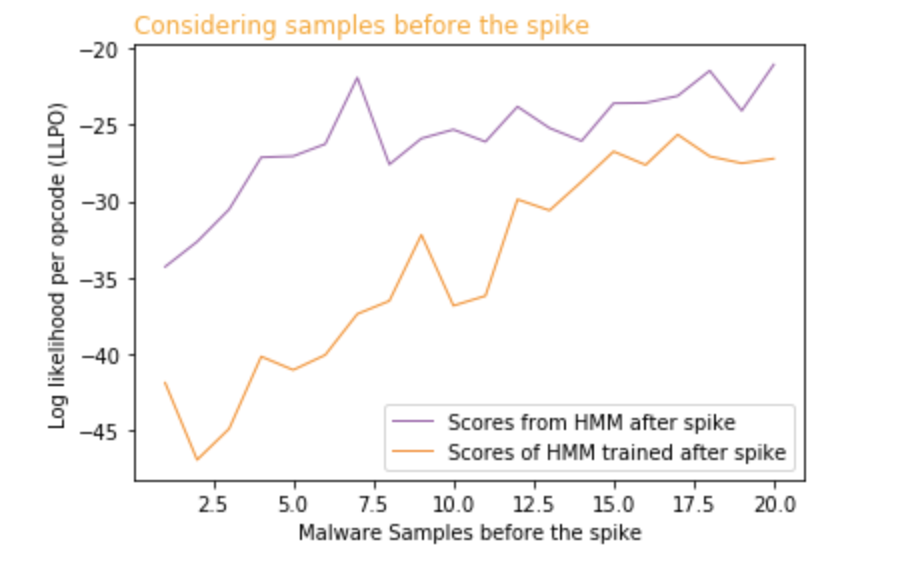}	
	& &	
	\includegraphics[width=0.35\textwidth]{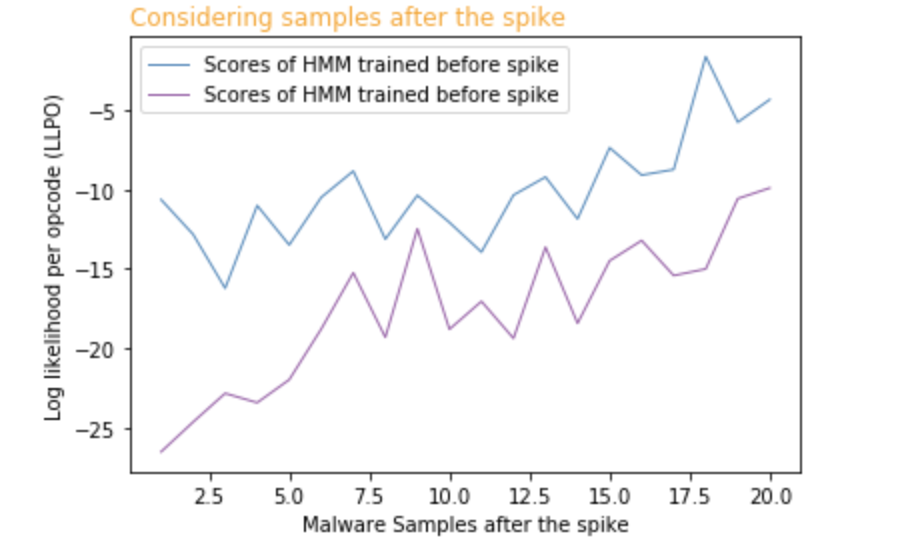}
	\\
	(a) BHO before spike
	& &
	(b) BHO after spike
	\\
	\includegraphics[width=0.35\textwidth]{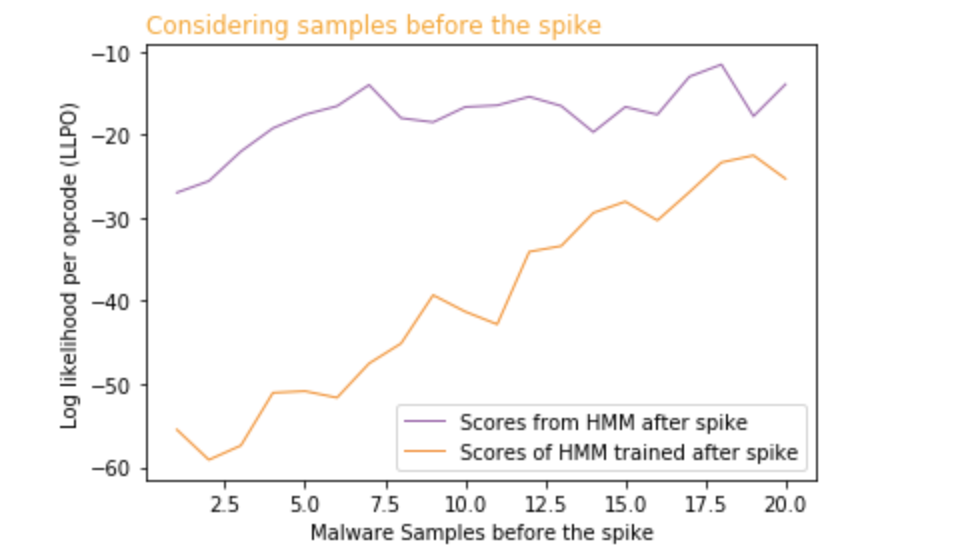}
	& &
	\includegraphics[width=0.35\textwidth]{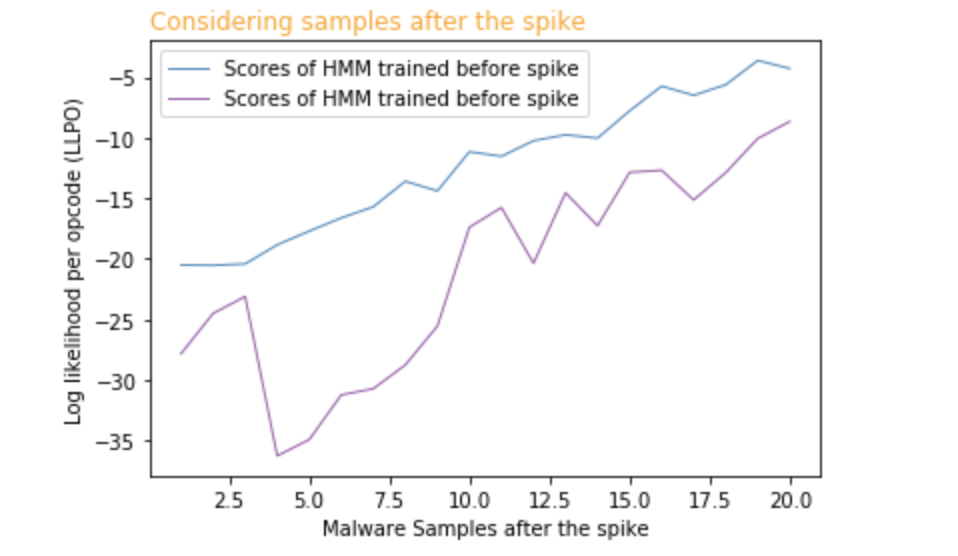}
	\\
	(c) Vobfus before spike
	& &
	(d) Vobfus after spike
	\end{tabular}
    \caption{HMM secondary tests for opcode-Word2Vec-SVM showing evolution}\label{fig:19}
\end{figure}

\begin{figure}[!htb]
	\centering
	\begin{tabular}{ccc}
	\includegraphics[width=0.325\textwidth]{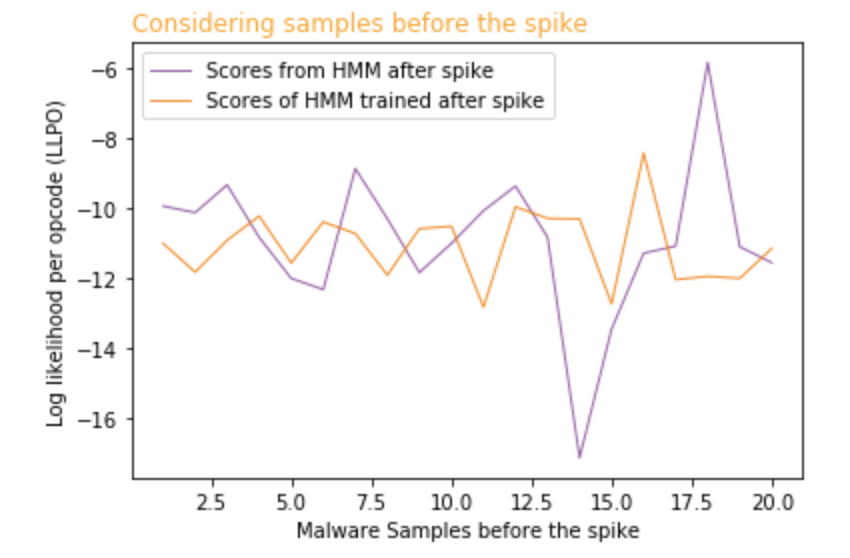}
	& &
	\includegraphics[width=0.35\textwidth]{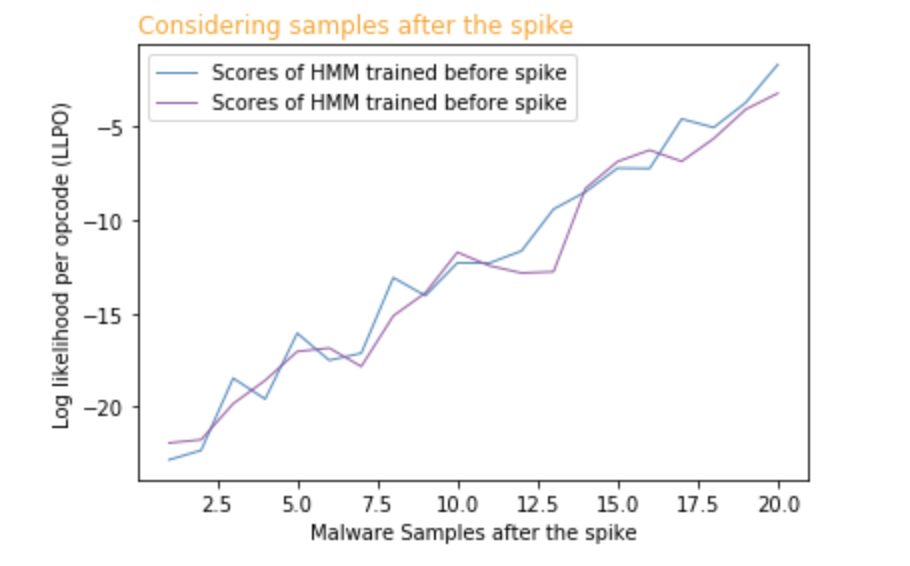}
	\\
	(a) CeeInject before spike
	& &
	(b) CeeInject after spike
	\end{tabular}
    \caption{HMM secondary tests for opcode-Word2Vec-SVM showing no evolution}\label{fig:19b}
\end{figure}

For the remaining four families in Figure~\ref{fig:13}, namely, 
CeeInject, Dorkbot, Hupigon, and Rbot, we do not detect
any significant spikes in their~$\chi^2$ graphs. 
Additional secondary HMM tests showing evolution
appear in Figure~\ref{fig:19}, while Figure~\ref{fig:19b} gives
an example of a secondary test that shows no evolution. 
It is evident that the results of these opcode-Word2Vec-SVM experiments are
a major improvement over the experiments considered above.

Of course, it is possible that there is no evolution to be detected, 
in some of these families. But, over the extended time periods under consideration, 
we believe it likely that evolution has occurred, which suggests
that the Word2Vec features are simply not sufficiently 
sensitive to detect changes in all cases.
In the next section, we consider another word embedding technique.
 
\subsection{Opcode-HMM2Vec-SVM Results}

The experiments in this section are essentially the same as those in the previous section, 
except that we use HMM2Vec embeddings in place of Word2Vec embeddings.
As discussed above, HMM2Vec embeddings use obtained from the~$B$
matrix of a trained HMM. 

\begin{figure}[!htb]
   \centering
   \begin{tabular}{ccc}
    \includegraphics[width=0.45\textwidth]{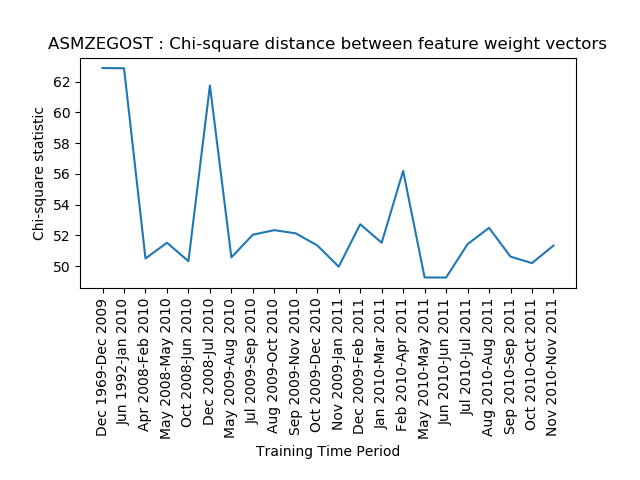}
    &
    &
    \includegraphics[width=0.45\textwidth]{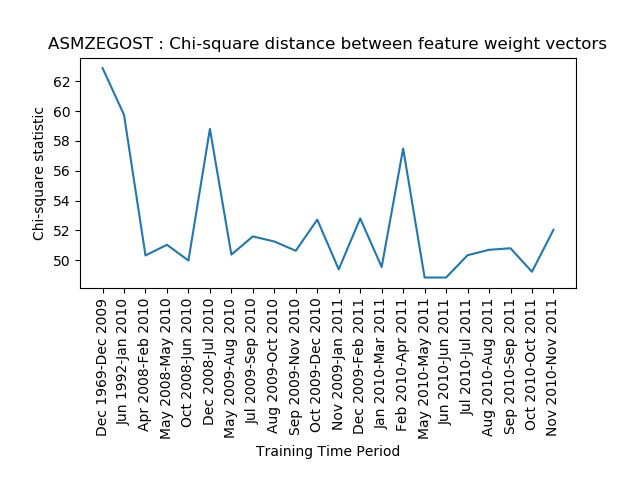}
    \\
    (a) Zegost without restarts
    &
    &
    (b) Zegost with~10 restarts
    \end{tabular}
    \caption{Opcode-HMM2Vec-SVM~$\chi^2$ graphs for Zegost}\label{fig:14}
\end{figure}

Figures~\ref{fig:14}~(a) and~(b) give HMM2Vec based~$\chi^2$
graphs for Zegost, using one random start and~10 random restarts, respectively.
Note that these results are based on HMMs with~$N=2$ hidden states,
which gives us embedding vectors of length~2.
Since our HMM training algorithm is a hill climb technique, multiple 
random restarts often enable us to find a stronger model. For the
Zegost results in Figure~\ref{fig:14}, random restarts appear to
offer little, if any, advantage. Consequently, for the remaining 
experiments in this section, we train a single HMM model, and we
do not perform any random restarts. 

We experimented with~$N=2$
and~$N=3$ hidden states (giving us embedding vectors of length~2 and~3,
respectively), but we did not find any improvement using~$N=3$.
Hence, we use HMM2Vec embedding vectors of length~2 in all experiments
below.

\begin{figure}[!htb]
   \centering
   \begin{tabular}{ccc}
    \includegraphics[width=0.33\textwidth]{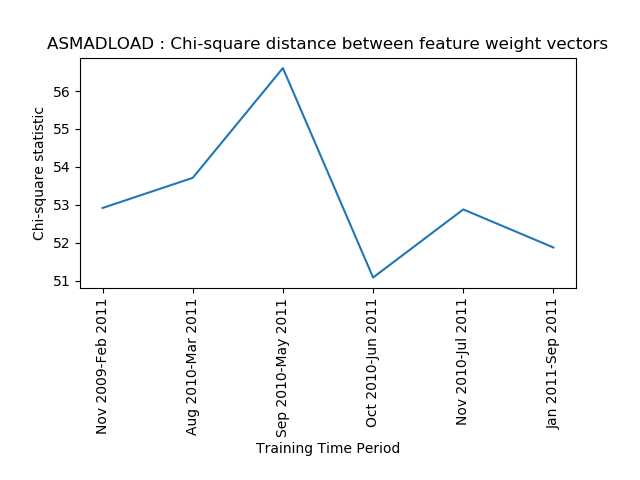}
    & 
    \includegraphics[width=0.33\textwidth]{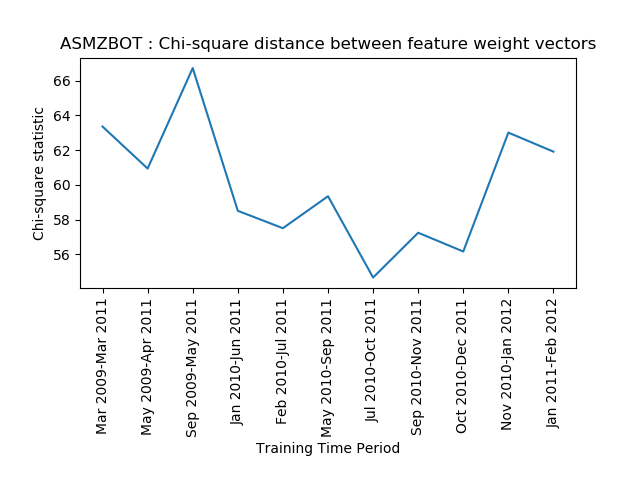}
    &
    \includegraphics[width=0.33\textwidth]{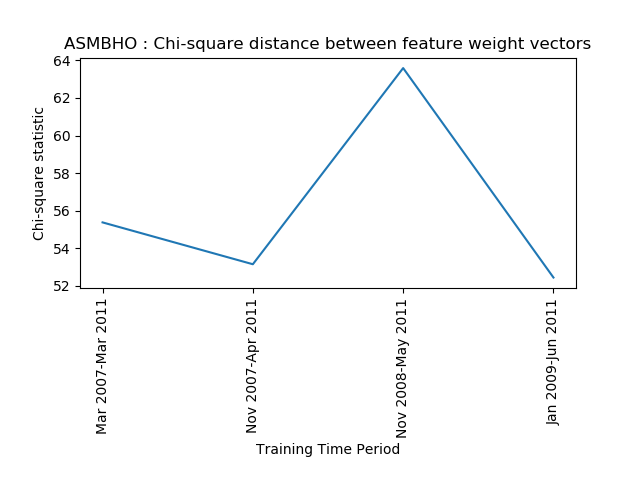}
    \\
    (a) Adload
    &
    (b) Zbot
    & 
    (c) BHO
    \\
    \includegraphics[width=0.33\textwidth]{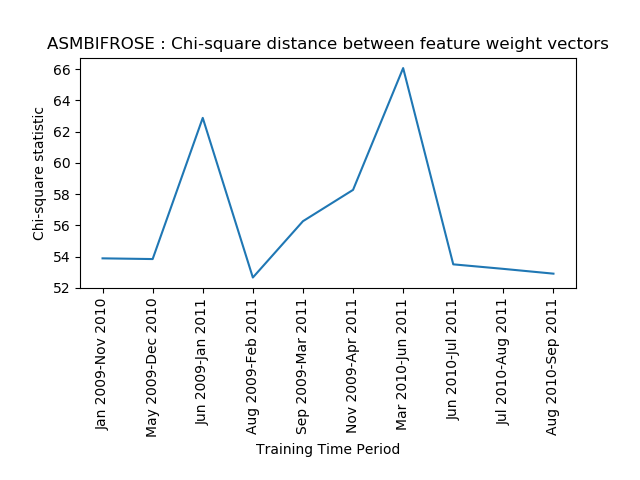}
    & 
    \includegraphics[width=0.33\textwidth]{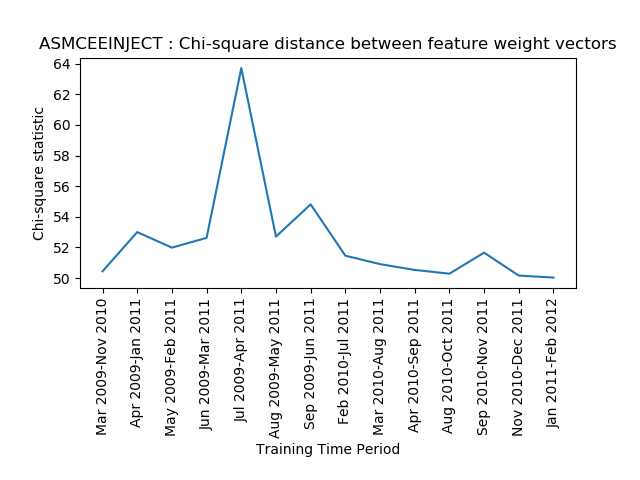}
    &
    \includegraphics[width=0.33\textwidth]{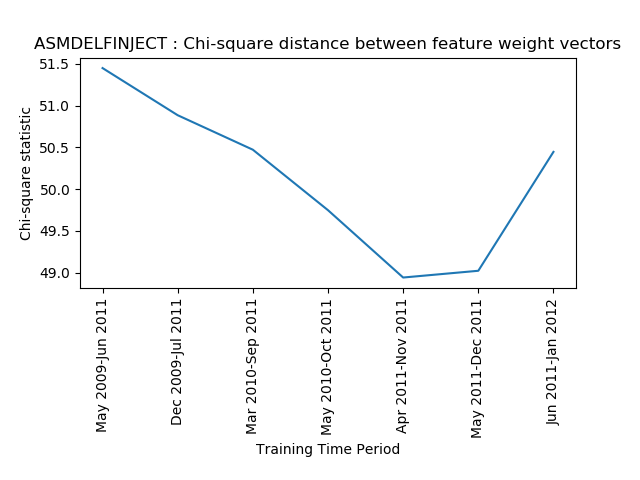}
    \\
    (d) Bifrose
    &
    (e) CeeInject
    &  
    (f) DelfInject
    \\[3ex]
    \multicolumn{3}{c}{
    \begin{tabular}{cc}
    \includegraphics[width=0.33\textwidth]{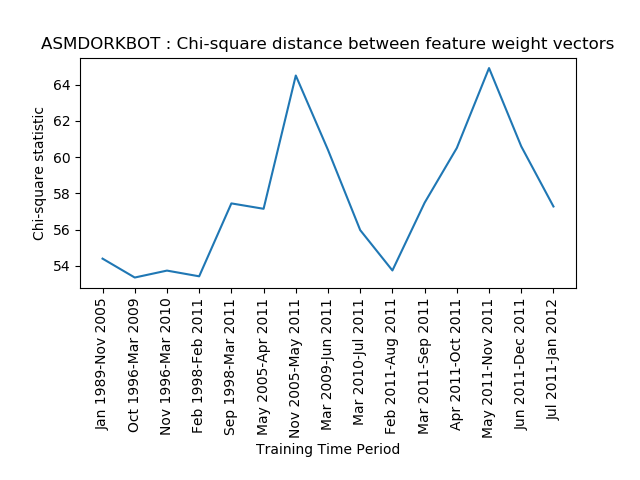}
    &
    \includegraphics[width=0.33\textwidth]{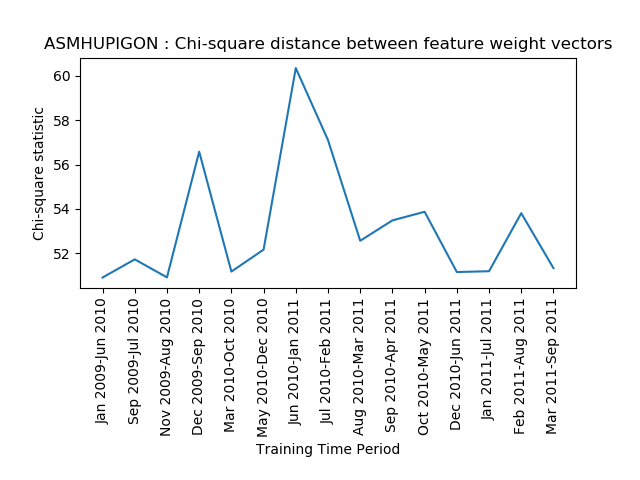}
    \\
    (g) Dorkbot
    & 
    (h) Hupigon
    \end{tabular}}
    \end{tabular}
    \caption{Opcode-HMM2Vec-SVM~$\chi^2$ for selected families}\label{fig:15}
\end{figure}

Figure~\ref{fig:15} shows the~$\chi^2$ graphs for~8 additional families
based on HMM2Vec embedding vectors. 
Overall, this HMM2Vec-SVM approach seems to provide better
results than the Word2Vec-SVM technique in the previous section, 
as we can detect more malware evolution using the HMM2Vec feature
engineering.

Based on the graphs in Figure~\ref{fig:15}, 
we make the following observations.
\begin{description}
    \item[\bf Adload]\hspace*{-8pt}--- An evolutionary event takes place in the time window Sep 2010--May 2011.
    \item[\bf Zbot]\hspace*{-8pt}--- No significant spike is observed in the~$\chi^2$ graph.
    \item[\bf BHO]\hspace*{-8pt}--- Significant evolution takes place in the time window Nov 2008--May 2011
    \item[\bf Bifrose]\hspace*{-8pt}--- Malware evolution takes place in the time window 
    March 2010--June 2011. The other spike in the period June 2009--Jan 2011 is a part of noise 
    in the data. This was confirmed by training HMMs on both sides. 
    \item[\bf CeeInject]\hspace*{-8pt}--- Evolution occurs in the time window June 2009--April 2011. 
    \item[\bf DelfInject]\hspace*{-8pt}--- No significant spike is observed in the~$\chi^2$ graph.
    \item[\bf Dorkbot]\hspace*{-8pt}--- No significant spike is observed in the~$\chi^2$ graph.
    \item[\bf Hupigon]\hspace*{-8pt}--- A significant spike in the time period June 2010--Jan2011 
    is observed in the graph.
\end{description}

\begin{figure}[!htb]
	\centering
	\begin{tabular}{ccc}
	\includegraphics[width=0.35\textwidth]{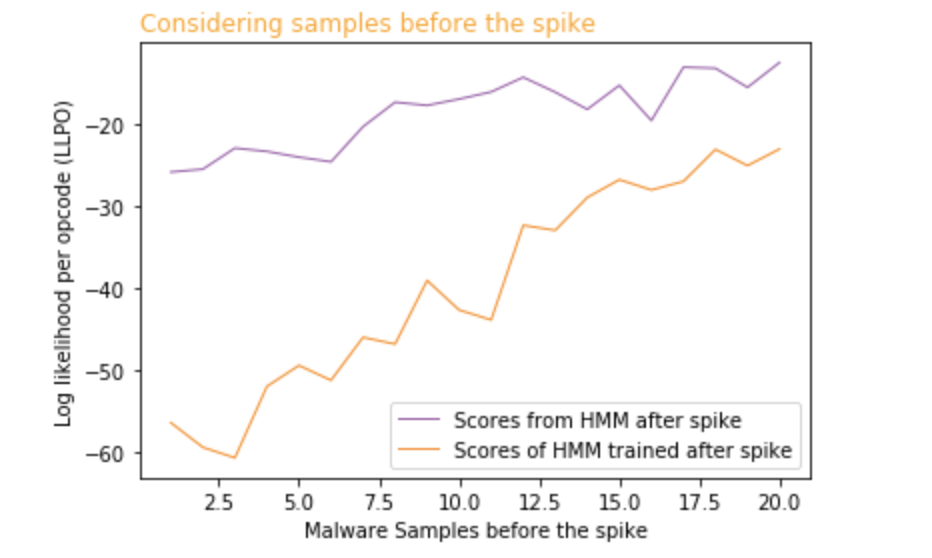}
	& &
	\includegraphics[width=0.35\textwidth]{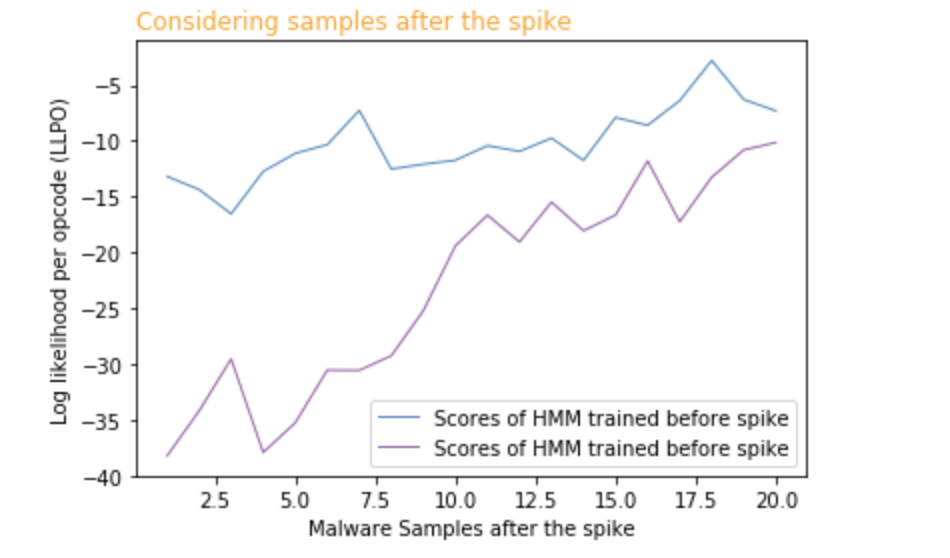}
	\\
	(a) BHO before spike
	& &
	(b) BHO after spike
	\\
	\includegraphics[width=0.35\textwidth]{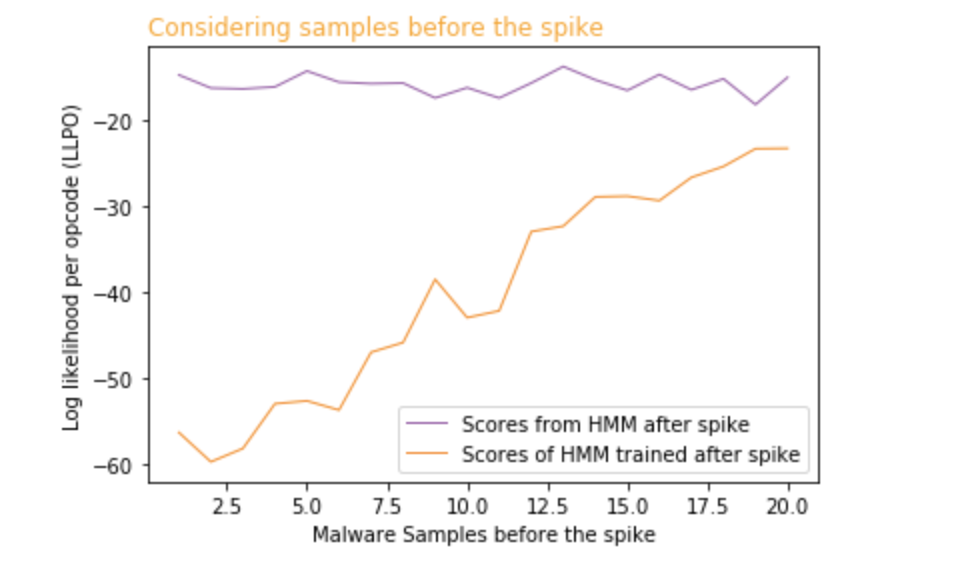}
	& &
	\includegraphics[width=0.35\textwidth]{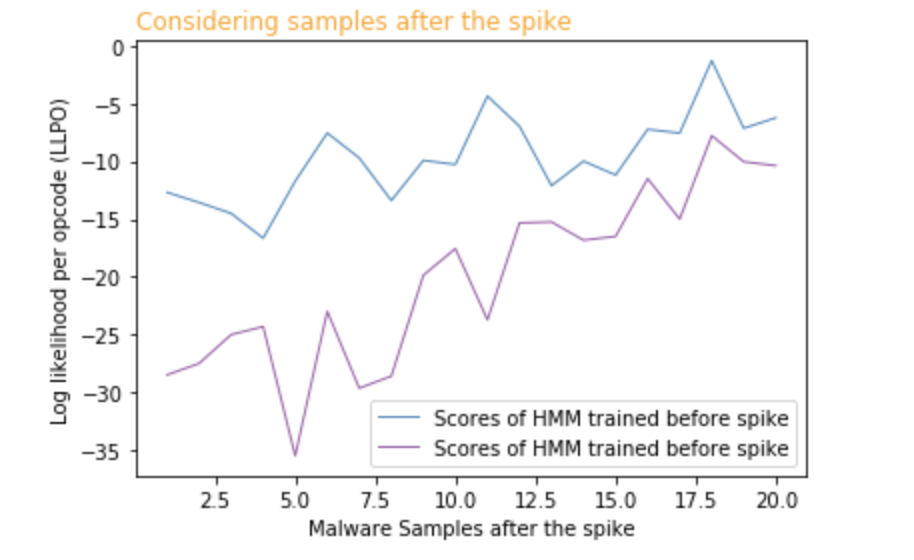}
	\\
	(c) Bifrose before spike
	& &
	(d) Bifrose after spike
	\\
	\includegraphics[width=0.35\textwidth]{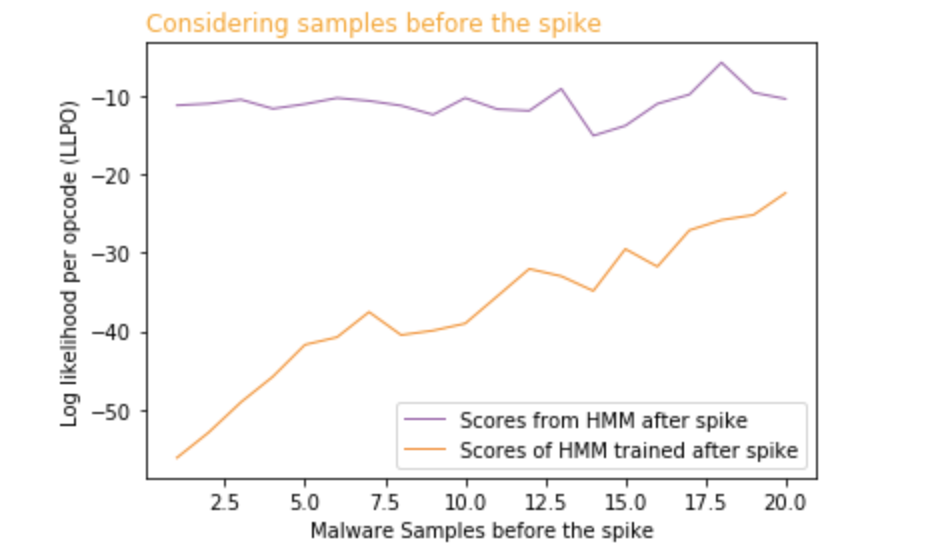}
	& &
	\includegraphics[width=0.35\textwidth]{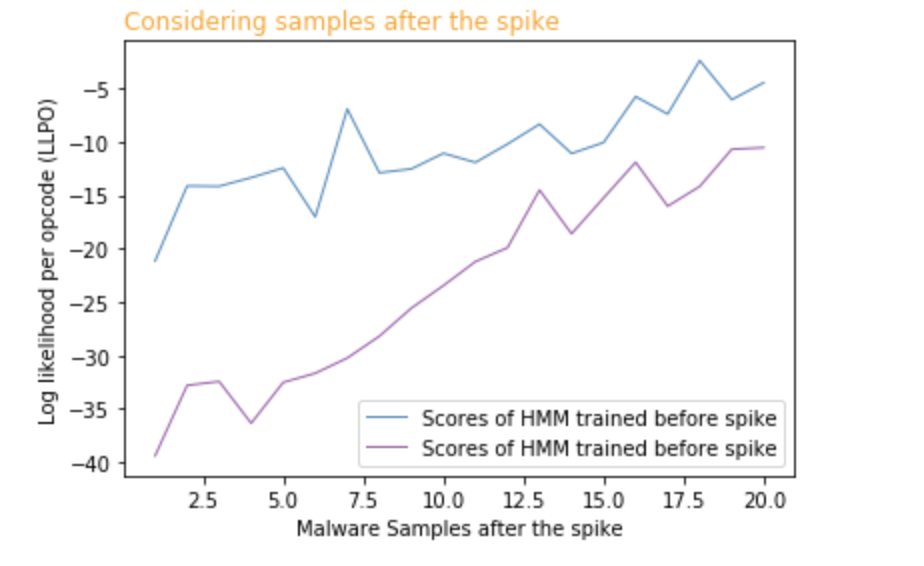}
	\\
	(e) CeeInject before spike
	& &
	(f) CeeInject after spike
	\\
	\includegraphics[width=0.34\textwidth]{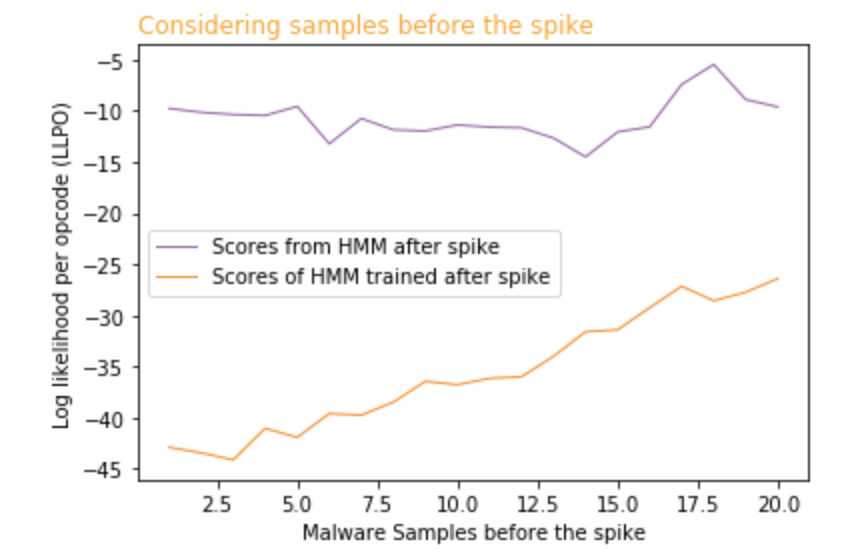}
	& &
	\includegraphics[width=0.36\textwidth]{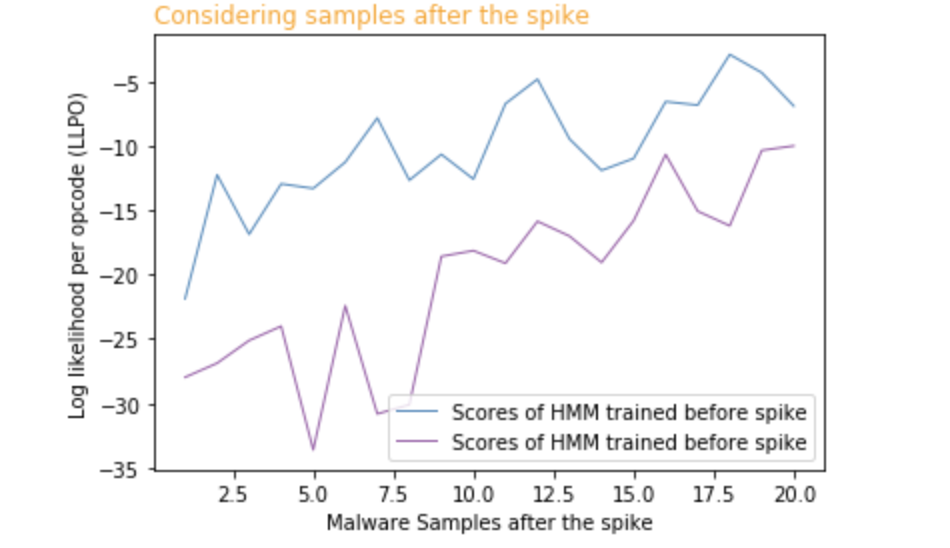}
	\\
	(g) Hupigon before spike
	& &
	(h) Hupigon after spike
	\end{tabular}
	\caption{HMM secondary tests for opcode-HMM2Vec-SVM showing evolution}\label{fig:23}
\end{figure}

\begin{figure}[!htb]
	\centering
	\begin{tabular}{ccc}
	\includegraphics[width=0.35\textwidth]{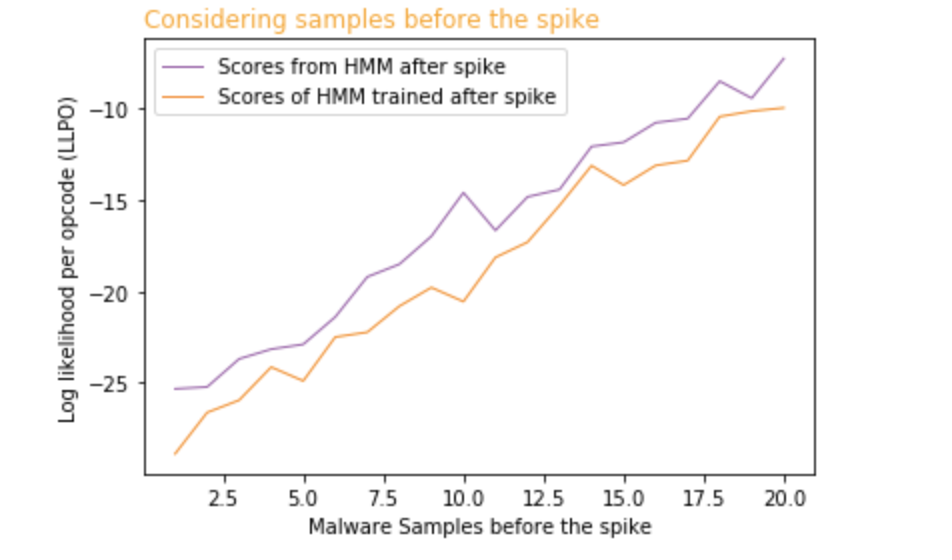}
	& &
	\includegraphics[width=0.37\textwidth]{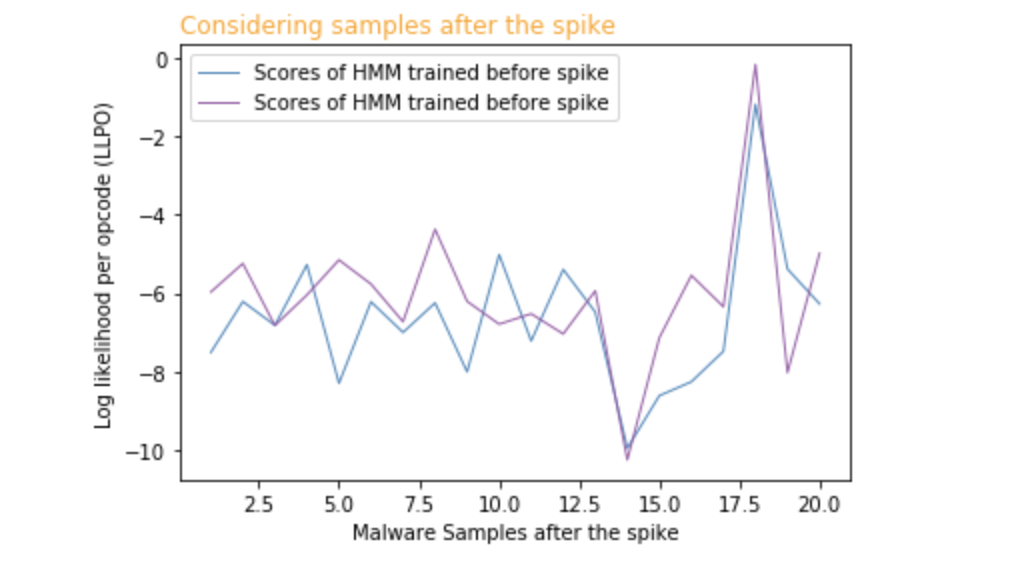}
	\\
	(a) DelfInject before dip
	& &
	(b) DelfInject after dip
	\\
	\includegraphics[width=0.315\textwidth]{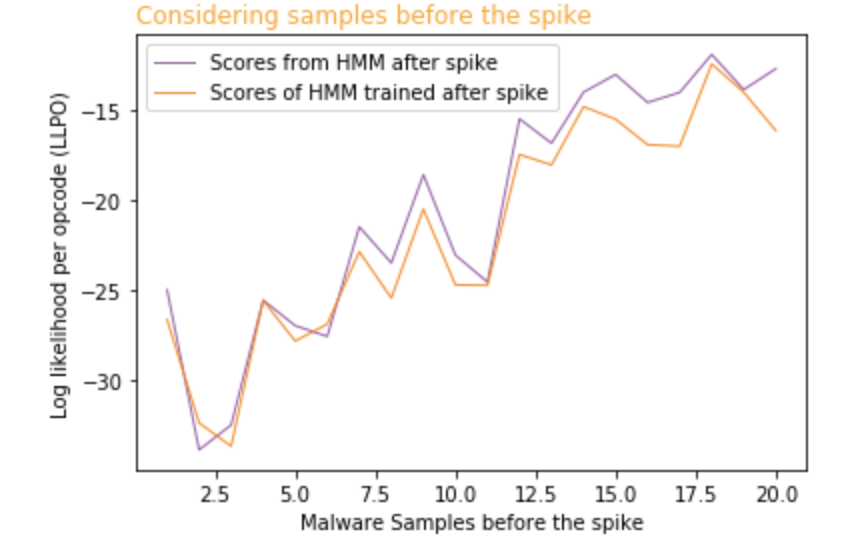}
	& &
	\includegraphics[width=0.37\textwidth]{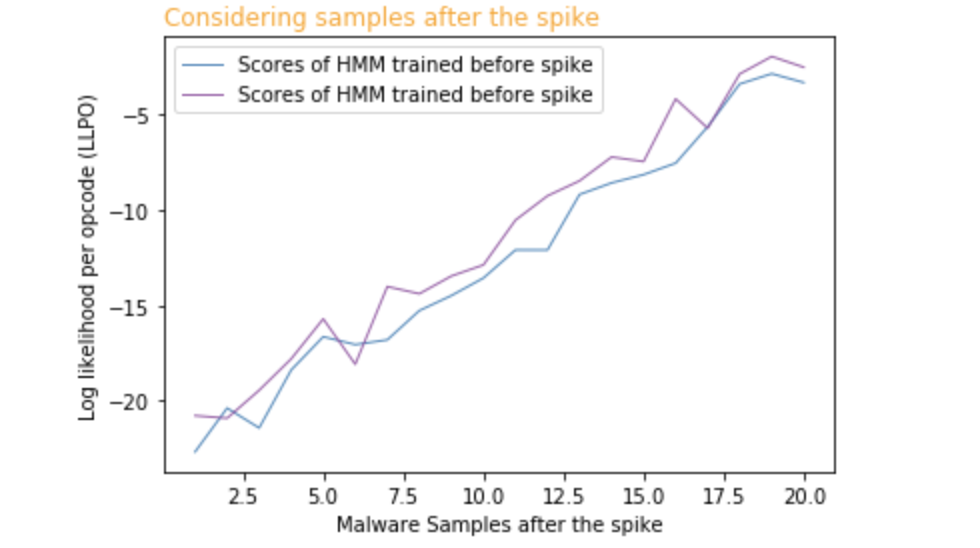}
	\\
	(c) Dorkbot before first spike
	& &
	(d) Dorkbot after first spike
	\end{tabular}
	\caption{HMM secondary tests for opcode-HMM2Vec-SVM showing no evolution}\label{fig:25}
\end{figure}

From Figure~\ref{fig:15} we conclude that we can observe significant spikes 
in almost all families using HMM2Vec-SVM analysis. 
For the families Adload, BHO, Bifrose, CeeInject, and Hupigon 
we observe significant spikes in the~$\chi^2$ distribution graph.
HMM secondary test confirming evolution for some of these cases
appear in Figure~\ref{fig:23}. In Figure~\ref{fig:25}, we give results
of HMM secondary tests that do not reveal evolution.

Comparing the families in which we could detect evolutionary changes 
using Word2Vec with those detected using HMM2Vec, we observe 
that the evolutionary points obtained using Word2Vec are also
found using HMM2Vec.
Yet, the HMM2Vec technique provides additional evolutionary points,
indicating that it is more sensitive to change than Word2Vec embeddings.
Overall, HMM2Vec performed better than any of the other approaches 
that we considered in this paper.


\section{Conclusion and Future Work}\label{chap:conclusion}

Previous research has shown that analysis based on PE file features and linear SVM 
models can be useful in detecting malware evolution~\cite{WadkarMayuri2019MMEU}.
In this paper, we expanded on---and improved upon---this previous work
in several ways. First, we considered opcode features, rather than PE file
features. Our intuition was that opcode based features would be more 
sensitive to the types of changes that we would like to detect, and our
results support this intuition. Second, we experimented with various
feature engineering techniques, and we found that vector embeddings
increase the sensitivity of the SVM analysis. Thirdly, we showed that a
secondary test using HMM techniques can be used to verify that
suspected evolutionary points in the timeline.
 
We experimented with a variety of techniques, and our best results 
were obtained using an approach that we refer to as opcode-HMM2Vec-SVM.
In this technique, we use mnemonic opcodes as raw features, then generate HMM2Vec
encodings of the opcodes, which serve as features for linear SVMs,
with the SVMs trained over sliding windows of time.
The resulting SVM weights are compared using a~$\chi^2$ statistic,
and we graph this statistic over the available timeline. Spikes in the~$\chi^2$
graph serve as indicators of likely evolutionary change. We were then
able to further confirm evolutionary changes using a secondary test
based on training HMMs on either side of a spike. 
This overall approach was more sensitive than previous work,
in the sense that we were able to detect additional changes
in the codebase of various families, and it was more precise,
since we have a secondary test available to confirm (or deny)
putative evolutionary changes.

In the realm of future work, additional machine learning techniques and additional
features and engineering strategies could be considered. For example,
neural network techniques could be used in place of SVMs,
with multiclass output probabilities playing the role of the linear SVM weights.
Another option would be to elevate our HMM-based secondary test 
to the role of the primary test. This might enable a more fine-grained
analysis of the timeline, as relatively little data is needed for HMM training.
With respect to feature engineering, dimensionality reduction techniques
would be a natural topic to consider.
The use of dynamic features might add value as well,
although the additional complexity involved with collecting such features
might be a concern.

\bibliographystyle{plain}

\bibliography{references.bib}

\end{document}